\documentclass[a4paper,12pt]{iopart}

\usepackage{iopams} 
\usepackage{amssymb}
\usepackage{epsfig,newlfont}
\usepackage{subfigure}
\usepackage[colorlinks, 
linkcolor=blue,
urlcolor=blue,
citecolor=blue,
bookmarksopen=false]{hype rref}
\usepackage{graphicx, xcolor}
\usepackage{caption}
\usepackage{epstopdf}

\newcommand{\bB}{\mathbf{B}}

\newcommand{\ivan}[1]{\textcolor{black}{#1}}
\newcommand{\jlvg}[1]{\textcolor{black}{#1}}

\def\newblock{\hskip .11em plus .33em minus .07em}

\begin{document}

\title{Large tangential electric fields in plasmas close to temperature screening}

\author{J. L.~Velasco$^1$, I. Calvo$^1$, J. M. Garc\'ia-Rega\~na$^1$, F. I. Parra$^{2,3}$, S. Satake$^{4,5}$, J. A. Alonso$^1$, {and the LHD team}}

\address{$^1$ Laboratorio Nacional de Fusi\'on, CIEMAT, Madrid, Spain}
\address{$^2$ Rudolf Peierls Centre for Theoretical Physics, University of Oxford, UK}
\address{$^3$ Culham Centre for Fusion Energy, Abingdon, UK}
\address{$^4$ National Institute for Fusion Science, Toki, Japan}
\address{\ivan{$^5$ The} Graduate School for Advanced Studies (SOKENDAI)}

\ead{joseluis.velasco@ciemat.es}

\begin{abstract}

Low-collisionality stellarator plasmas usually display a large negative radial electric field that has been expected to cause accumulation of impurities due to their high charge number. In this paper, two combined effects that can potentially modify this scenario are discussed. First, it is shown that, in low collisionality plasmas, the kinetic contribution of the electrons to the radial electric field can make it negative but small, bringing the plasma close to impurity temperature screening (i.e., to a situation in which the ion temperature gradient is the main drive of impurity transport and causes outward flux); in plasmas of very low collisionality, such as those of the Large Helical Device displaying impurity hole~\cite{ida2009observation,yoshinuma2009observation}, screening may actually occur. Second, the component of the electric field that is tangent to the flux surface (in other words, the variation of the electrostatic potential on the flux surface), although smaller than the radial component, has recently been suggested to be an additional relevant drive \ivan{for} radial impurity transport. Here, it is explained that, especially when the radial electric field is small, the tangential magnetic drift has to be kept in order to correctly compute the tangential electric field, that can be larger than previously expected. This can have a strong impact on impurity transport, as we illustrate by means of simulations using the newly-developed code \texttt{KNOSOS} (KiNetic Orbit-averaging-SOlver for Stellarators).

\end{abstract}

\section{Introduction}\label{SEC_INTRO}

Achieving impurity control is a crucial issue in the path towards a fusion reactor based on magnetic confinement. This is a problem that is \ivan{especially} relevant for stellarators, whose standard high temperature operation scenarios typically display impurity accumulation. There are few known exceptions: the High Density H-mode, observed at the helias Wendelstein 7-AS at very high densities~\cite{mccormick2002hdh}, and the impurity hole, observed at the Large Helical Device (LHD) at very low collisionalities~\cite{ida2009observation,yoshinuma2009observation}. Since temperature screening (i.e., outward impurity flux driven by the ion temperature gradient) is not expected for stellarators from basic considerations (see e.g.~\cite{igitkhanov2006impurity}) unlike in tokamaks, there has been much theoretical work~\cite{regana2013euterpe,regana2017phi1,mikkelsen2014gk,nunami2016iaea,alonso2016inertia,helander2017prl} devoted to try to understand these exceptions.

The lack of temperature screening in typical stellarator scenarios is caused by the fact that both the radial electric field $E_r$ and the ion temperature gradient $T_i'$ contribute to radial impurity transport; the latter, through an outward pinch roughly given by $T_i'$; the former, by means of and inward pinch $Z_IeE_r$ (\ivan{$Z_Ie$ being} the charge of the impurity and $e$ the proton charge). Even for peaked ion temperature profiles, in standard ion-root conditions~\cite{dinklage2013ncval} the radial electric field is bound to be as large as the ion temperature gradient, $eE_r\sim T_i'$, and $Z_I\gg 1$ ensures impurity accumulation. 

The large charge number $Z_I$ anticipates one of the additional physical mechanisms that is a candidate for explaining how impurities are flushed out from the plasma despite the negative radial electric field (that is, the exceptions mentioned above): the electrostatic potential is only approximately a constant on the flux surfaces of a stellarator~\cite{regana2013euterpe,regana2017phi1,calvo2017sqrtnu} and, for low collisionalities of the bulk species, its variation on the flux surface, that we denote by $\varphi_1$ (in other words, the component of the electric field that is tangent to the flux surface), can in principle be a relevant drive \ivan{for} the radial transport of moderate-to-high $Z_I$ impurities. However, the simulations done so far~\cite{regana2013euterpe,regana2017phi1} do not generally predict an outwards-directed impurity flux like the one \ivan{observed, for example, in the impurity hole}: the $Z_Ie\varphi_1$ term brings the impurity flux closer to zero in some situations, but its effect is not large enough.

In this paper we show that for low-collisionality plasmas of stellarators the radial electric field can actually be smaller than what the above discussion suggests, and that, on the contrary, the tangential electric field is larger than what previous numerical simulations have predicted, and both effects have a large impact on the impurity flux. First, as we will discuss throughout the next sections, the standard ion root description neglects the contribution of the electrons to the ambipolarity equation. Once included, $E_r$ is still negative but gets smaller in magnitude, and then $T_i'$ is able to compete with it in driving impurity transport; for extreme cases, screening may occur, as first discussed in~\cite{velasco2017hole}. Second, previous calculations of $\varphi_1$~\cite{regana2013euterpe,regana2017phi1} neglect the tangential magnetic drift in the drift-kinetic equation, and therefore the contribution of the superbanana-plateau regime to bulk ion transport and the contribution of the superbanana-plateau layer to the variation of the ion bulk density and of the electrostatic potential on the flux surface. These contributions may become specially relevant for plasmas of small radial electric field, as shown recently in reference~\cite{calvo2017sqrtnu}, which derives for the first time radially-local equations for the calculation of radial transport and $\varphi_1$ including the tangential magnetic drift (and valid, as we will discuss, for stellarators close to omnigeneity, where the non-omnigeneous perturbation has small gradients). In this paper, we illustrate these two effects by means of simulations using \texttt{DKES}~\cite{hirshman1986dkes}, \texttt{EUTERPE}~\cite{regana2017phi1}  and the newly-developed code \texttt{KNOSOS} (KiNetic Orbit-averaging-SOlver for Stellarators)\jlvg{, which solve different versions of the \ivan{drift kinetic equation} that will be presented in the next sections.}

The rest of the paper is organized as follows. In section~\ref{SEC_SCREENING}, we discuss impurity screening in stellarators at low collisionalities without including the effect of $\varphi_1$. The equations employed are outlined in \ivan{subsection~\ref{SEC_EQUATIONS1}. They} correspond to the most standard neoclassical models, and we put the emphasis on the terms that, associated to the kinetic contribution of electrons and typically neglected in the discussion of the ambipolarity condition, become relevant at low collisionalities. Then, in subsection~\ref{SEC_EXAMPLES}, we present some examples of real plasmas of LHD that illustrate the physics discussed in the previous \ivan{subsection. It} will be shown that plasmas of low collisionality are not far from impurity screening, since for low enough collisionalities the radial electric field is negative but small in size, and the temperature gradient is then able to compete with it in driving impurity transport. But this balance is not enough to accurately describe plasmas such as those displaying impurity hole, and additional physical effects are needed. This will lead to the second part of the paper. In section~\ref{SEC_LARGEEPAR}, we evaluate the influence of the tangential magnetic \ivan{drift on} the variation of the electrostatic potential on the flux surface and on impurity transport. The model used for the calculation of $\varphi_1$ is outlined in subsection~\ref{SEC_EQUATIONS2}, and a newly developed code, \texttt{KNOSOS}, able to solve the equations of the model will be briefly presented in subsection~\ref{SEC_KNOSOS} (and benchmarked against \texttt{EUTERPE}, a reference neoclassical code for the calculation of the tangential electric field~\cite{regana2017phi1}). In subsection~\ref{SEC_RESULTS}, we will apply this model to low collisionality LHD plasmas and we will present the first \ivan{results. We} will see that the tangential electric field becomes larger than previously expected from standard neoclassical models in which the tangential magnetic drift is neglected, and that this can have a very important impact on impurity transport. The conclusions will come in section~\ref{SEC_CONCLUSIONS}.

\section{Small radial electric field and temperature screening in \jlvg{standard neoclassical models}}\label{SEC_SCREENING}

\subsection{Equations}\label{SEC_EQUATIONS1}

Neoclassical transport of the bulk species in low-collisionality stellarator plasmas is typically solved in the \ivan{large} aspect ratio limit and assuming relatively large values of the radial electric field. In this limit, the relevant drift-kinetic equation reads
\begin{equation}
\overline{\mathbf{\ivan{v_{E,0}}}\cdot\nabla\alpha}~\partial_\alpha g_b +\overline{\mathbf{v_M}\cdot\nabla r}~F_{M_b} \Upsilon_b = \overline{C(g_b)}\,,\label{EQ_DKE}
\end{equation}
see, e.g.~\cite{beidler2011ICNTS,calvo2017sqrtnu}. In this equation, $g_b(r,\alpha,v,\lambda$) is, for species $b$, the dominant piece of the non-adiabatic component of the deviation of the  distribution function from a Maxwellian\jlvg{
\begin{equation}
F_{M_b} =n_b\left(\frac{\ivan{m_b}}{2\pi T_b}\right)^{3/2}\exp{\left(-\frac{\ivan{m_b}v^2}{2T_b}\right)}
\end{equation}
with density \ivan{$n_b(r)$ and temperature $T_b(r)$}.} \ivan{$C$ is the pitch-angle-scattering collision operator and $\mathbf{v_M}$ is the magnetic drift,
\begin{equation}
\mathbf{v_M} = \frac{m_b}{2Z_be}(v^2+v_\parallel^2)\frac{\mathbf{B}\times\nabla B}{B^3}\,,
\end{equation}
where $m_b$ and $Z_be$ are the charge and mass of species $b$. Here, $r$ is a radial coordinate (we choose it to be $r = a \sqrt{\Psi_t/\Psi_{t,LCMS}}$, where $\Psi_t$ is the toroidal magnetic flux over $2\pi$, $LCMS$ denotes the last-closed magnetic surface, and $a$ is the minor radius of the stellarator), prime denotes derivative with respect to $r$, $\alpha$ is a poloidal angle that labels magnectic field lines on the flux surface and $l$ is the arc length along the field line. In these coordinates, the magnetic field $\mathbf{B}$ reads
\begin{equation}
\bB = \Psi'_t(r) \nabla r\times \nabla\alpha.
\end{equation}
As for the velocity coordinates, $v$ is the magnitude of the velocity, $\lambda = v_\perp^2/(v^2B)$ is the pitch-angle coordinate and $v_\perp$ is the component of the velocity perpendicular to the magnetic field. The function $g_b$ is independent of $l$ and vanishes for passing trajectories, and the coefficients of the equation are averages over the trapped orbits, defined as
\begin{equation}\label{eq:orbit_average}
\overline{f}(r,\alpha,v,\lambda)= \ivan{\frac{2}{\tau}\int_{l_{b_1}}^{l_{b_2}}f(r,\alpha,l,v,\lambda)|v_\parallel|^{-1}\mathrm{d}l \,} ,
\end{equation}
where
\begin{equation}
|v_\parallel|(r,\alpha,l,v,\lambda) = v\sqrt{1-\lambda B(r,\alpha,l)} \, 
\end{equation}
is the magnitude of the parallel velocity (the average defined in (\ref{eq:orbit_average}) is correct for functions that are even in $v_\parallel$, which is all we need in this paper), $l_{b_1}$ and $l_{b_2}$ are the bounce points of the orbit, i.e. the solutions of $1-\lambda B = 0$, and
\begin{equation}
\tau= 2\int_{l_{b_1}}^{l_{b_2}}|v_\parallel|^{-1}\mathrm{d} l
\end{equation}
is the time that it takes for a particle to complete the orbit. 
}

\ivan{
We denote by $\mathbf{v_E}$ the $\mathbf{E}\times \mathbf{B}$ drift,
\begin{equation}
\mathbf{v_E} = \frac{\mathbf{B}\times\nabla\varphi}{B^2}\,.
\end{equation}
In \ivan{equation}~(\ref{EQ_DKE}), the electrostatic potential, $\varphi$, is assumed to be a flux function, $\varphi\approx \varphi_0(r)$, and $E_r\equiv -\varphi_0'$. On the left side of (\ref{EQ_DKE}), $\mathbf{v_{E,0}}$ is the $\mathbf{E}\times \mathbf{B}$ drift due to $\varphi_0$.
}

Finally, $\Upsilon_b$ is a combination of the thermodynamic \ivan{forces,}
\begin{equation}
\Upsilon_b = \frac{n_b'}{n_b} + \frac{T_b'}{T_b}\left(\frac{m_bv^2}{2T_b}-\frac{3}{2}\right)-\frac{Z_beE_r}{T_b}\,.
\end{equation}

In the trace-impurity limit, for given density $n_b$ and temperature $T_b$ profiles, the radial electric field can be calculated by imposing ambipolarity of the \ivan{neoclassical radial particle fluxes,}
\begin{equation}
Z_i\langle\mathbf{\Gamma}_i\cdot\nabla r\rangle = \langle\mathbf{\Gamma}_e\cdot\nabla r\rangle\,. \label{EQ_AMB}
\end{equation}
Here, $\langle ... \rangle$ denotes flux-surface average, and the particle flux can written using the solution of equation (\ref{EQ_DKE}) in\jlvg{
\begin{equation}
\langle\ivan{\mathbf{\Gamma}_b}\cdot\nabla r\rangle  =\frac{\ivan{\pi \Psi_t'}}{V'}\int_0^\infty\mathrm{d}v\int_{B^{-1}_{{\mathrm{max}}}}^{\ivan{B^{-1}_{{\mathrm{min}}}}}\mathrm{d}\lambda\,\int_0^{2\pi}\mathrm{d}\alpha\,v^3 \tau\,\overline{\mathbf{v_M}\cdot\nabla r}\,g_b\,\label{EQ_FLUX}
\end{equation}
where $V$ is the volume enclosed by the flux} \ivan{surface, and $B_{\mathrm{max}}$ and $B_{\mathrm{min}}$ are the maximum and minimum values of $B$ on the flux surface, respectively}.

It is straightforward to write the flux-surface-averaged radial flux of species $b$ for low collisionalities \ivan{as}
\begin{equation}
\frac{\langle\mathbf{\Gamma}_b\cdot\nabla r\rangle}{n_b}=- L_1^b\left(\frac{n_b'}{n_b}- \frac{Z_beE_r}{T_b}\right)-L_2^b\frac{T_b'}{T_b}\,,\label{EQ_FLUX2}
\end{equation}
\normalsize
where $L_1^b$ and $L_2^b$ are (positive and in general dependent of $n_b$, $T_b$ and $E_r$) neoclassical transport \jlvg{\ivan{coefficients,}
\begin{eqnarray}
\fl L_1^b  &=&-\frac{\ivan{\pi \Psi_t'}}{n_bV'}\int_0^\infty\mathrm{d}v\int_{B^{-1}_{{\mathrm{max}}}}^{\ivan{B^{-1}_{{\mathrm{min}}}}}\mathrm{d}\lambda\,\int_0^{2\pi}\mathrm{d}\alpha\,v^3 \tau\,\overline{\mathbf{v_M}\cdot\nabla r}\,\ivan{\frac{g_b}{\Upsilon_b}\,,}\nonumber\\
\fl L_2^b  &=&-\frac{\ivan{\pi \Psi_t'}}{n_bV'}\int_0^\infty\mathrm{d}v\int_{B^{-1}_{{\mathrm{max}}}}^{\ivan{B^{-1}_{{\mathrm{min}}}}}\mathrm{d}\lambda\,\int_0^{2\pi}\mathrm{d}\alpha\,v^3 \tau\,\overline{\mathbf{v_M}\cdot\nabla r}\,\left(\frac{m_bv^2}{2T_b}-\frac{3}{2}\right)\ivan{\frac{g_b}{\Upsilon_b}\,.}
\end{eqnarray}
} The calculation of these transport coefficients by solving equation~(\ref{EQ_DKE}) is the standard approach used by a large variety of neoclassical codes~\jlvg{such as DKES~\cite{hirshman1986dkes}}, see~\cite{beidler2011ICNTS} and references therein.

The radial electric field is set for given plasma profiles by ambipolarity of the neoclassical fluxes, equation~(\ref{EQ_AMB}). However, the discussion about impurity screening in stellarators is typically done in the so-called ion root approximation, where the electron contribution is neglected, and this amounts to obtaining the radial electric field from:
\begin{equation}
\langle\mathbf{\Gamma}_i\cdot\nabla r\rangle (E_r)=0\,,
\end{equation}
which can be rearranged as: 
\begin{equation}
\frac{eE_r}{T_i} = \frac{1}{Z_i}\frac{n_i'}{n_i} + \frac{1}{Z_i}\frac{L_2^i}{L_1^i}\frac{T_i'}{T_i} \approx \frac{5}{4}\frac{T_i'}{T_i}\,.\label{EQ_ERIROOT}
\end{equation}\normalsize
Here, we have assumed that the ion density profile is flat ($n_i'=0$), that the bulk ions are singly-charged ($Z_i=1$) and they are in the $\sqrt{\nu}$ regime $\left(\frac{L_2^i}{L_1^i} \approx\frac{5}{4}\right)$~\cite{maassberg1999densitycontrol}. We end up with a radial electric field that it is proportional to the ion temperature gradient\jlvg{, and such that $eE_r\!\approx\!T_i'$.}

Equation~(\ref{EQ_FLUX}) cannot be generally used as such for impurities $I$, as additional terms coming from the bulk species have to be taken into account (and may reduce the role of the radial electric field~\cite{helander2017prl}). However, it is consistently observed in simulations and in experiments that a negative (positive) radial electric field drives impurities inwards (outwards), so it is worth comparing the size of the \ivan{$Z_I e E_r$ term} with other terms driving impurity transport, such as $T_I'$. Using equation~(\ref{EQ_ERIROOT}) and asumming that $T_I=T_i$, we \ivan{obtain}
\begin{equation}
Z_I\frac{5}{4}\frac{|T_i'|}{T_i} \gg \frac{L_2^I}{L_1^I}\frac{|T_i'|}{T_i}\,.
\end{equation}
We see that, as soon as the impurity charge $Z_Ie$ is large enough, the term coming from the radial electric field becomes larger, since $L_2^I/L_1^I\sim 1$. 

\begin{figure}
\begin{center}
\includegraphics[width=\columnwidth,angle=0]{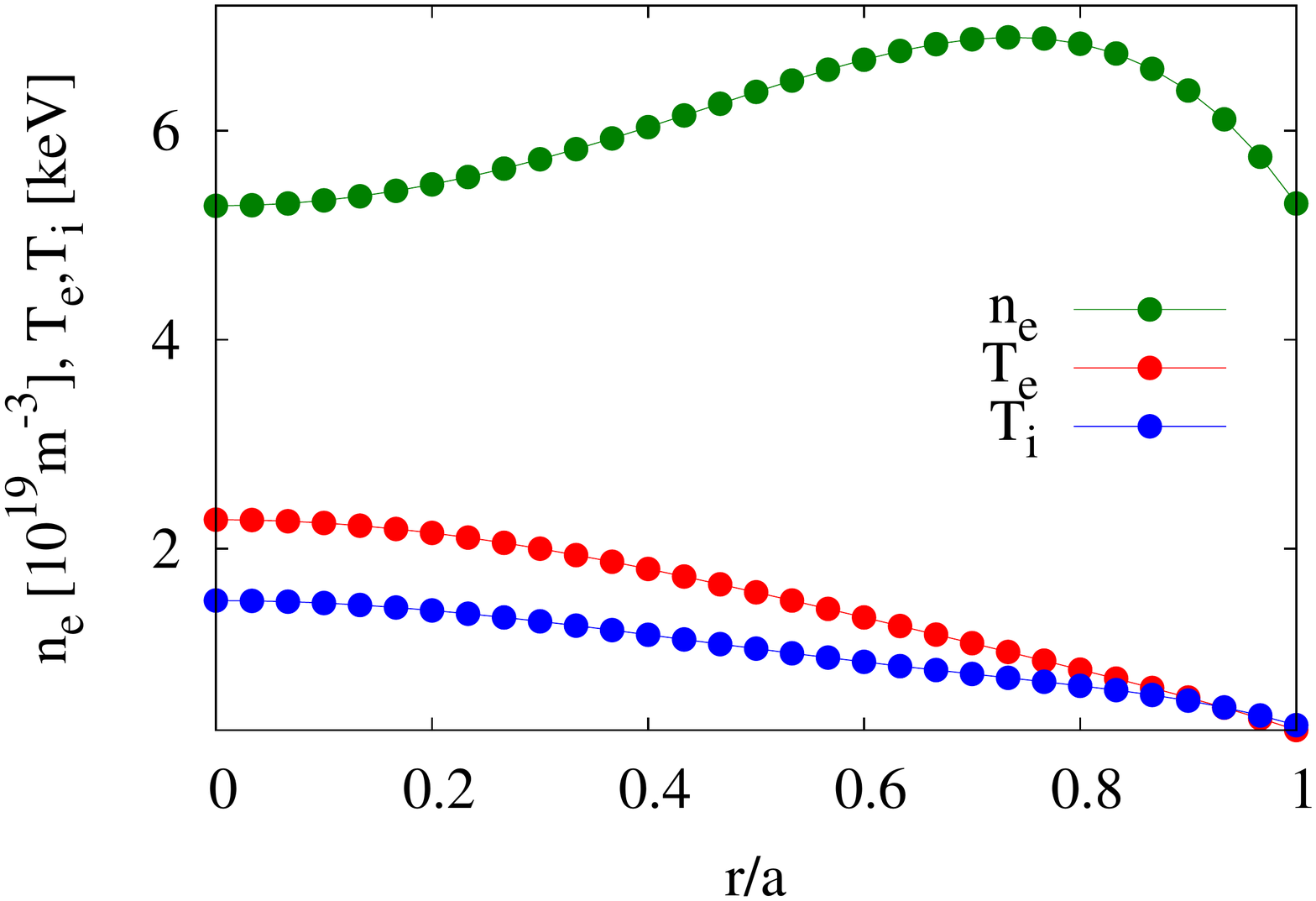}\vskip-2.5cm
\includegraphics[width=\columnwidth,angle=0]{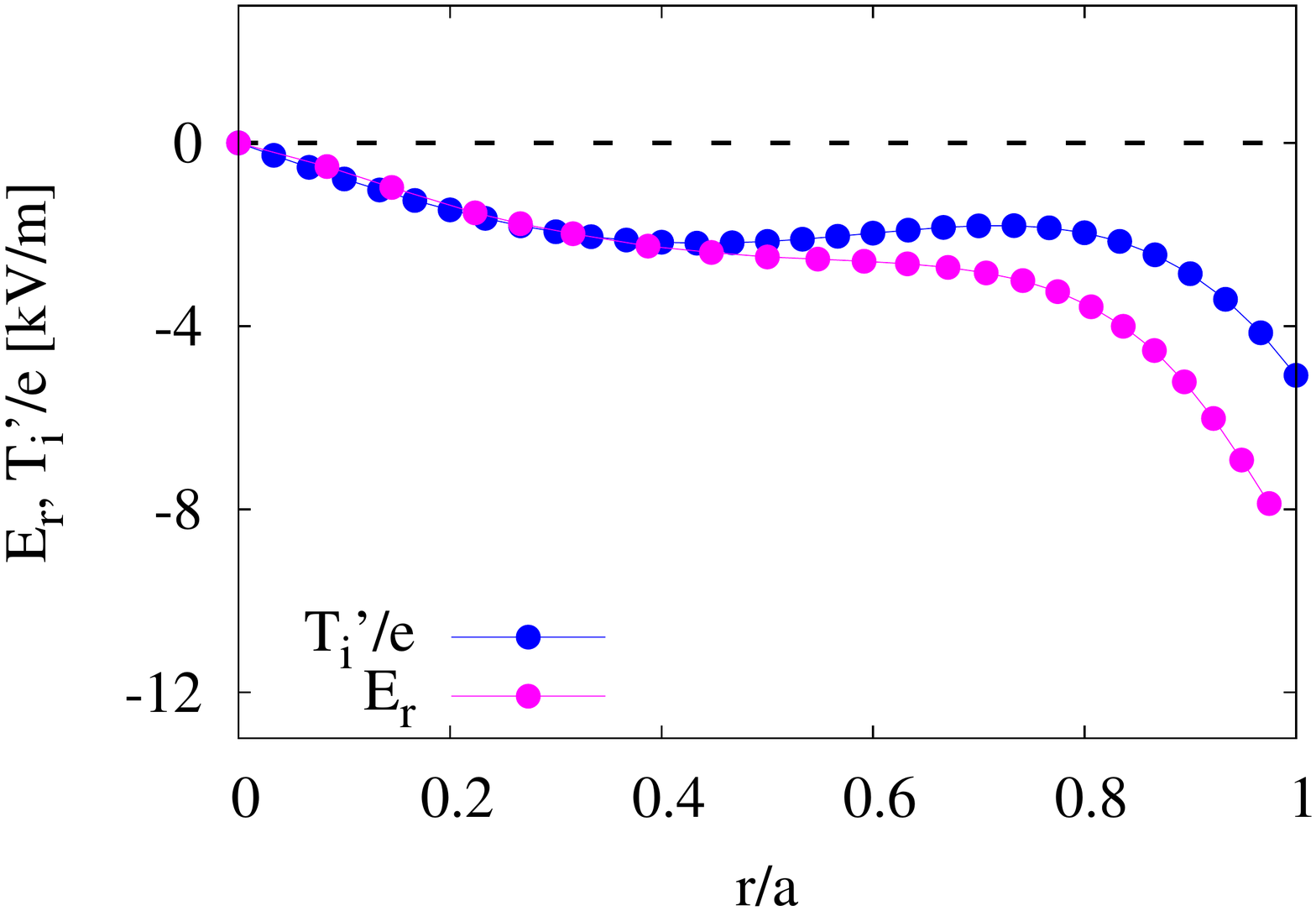}
\end{center}
\vskip-1cm\caption{Plasma profiles for an ion root plasma of LHD (top), and calculated radial electric field (bottom).}
\label{FIG_IR}
\end{figure}

At this point we remind the reader that the key reason for this result has been that the radial electric field is negative and comparable to the ion temperature gradient, as a consequence of the ion root approximation. While this is justified for some plasma parameters, \jlvg{we will see that, at low collisionalities, taking into account the electrons in \ivan{equation~(\ref{EQ_AMB}) changes the results significantly}}. Then, one can write a more general expression for the radial electric field, in which it becomes dependent also on the electron temperature gradient,
\begin{equation}
\frac{eE_r}{T_i} \approx \frac{L_2^i\frac{T_i'}{T_i} - {L_2^e\frac{T_e'}{T_e}}}{L_1^i+{\frac{T_i}{T_e}L_1^e}}\,.
\end{equation}
These new terms are typically negligible, due to the small Larmor radius of the electrons (compared to that of the ions) at similar temperatures but, in the low collisionality regimes of stellarators (specifically in the $1/\nu$ regime), the electron transport coefficients may grow very fast with the temperature and compete with that of the ions. Depending on their size, these additional terms may lead to a situation in which \jlvg{$eE_r$ is still negative, \ivan{but much} smaller than $T_i'$}. In this scenario, temperature screening for low-$Z_I$ impurities may happen. And, generally speaking, the closer to this situation, the closer to impurity screening.

\begin{figure}
\begin{center}
\includegraphics[width=\columnwidth,angle=0]{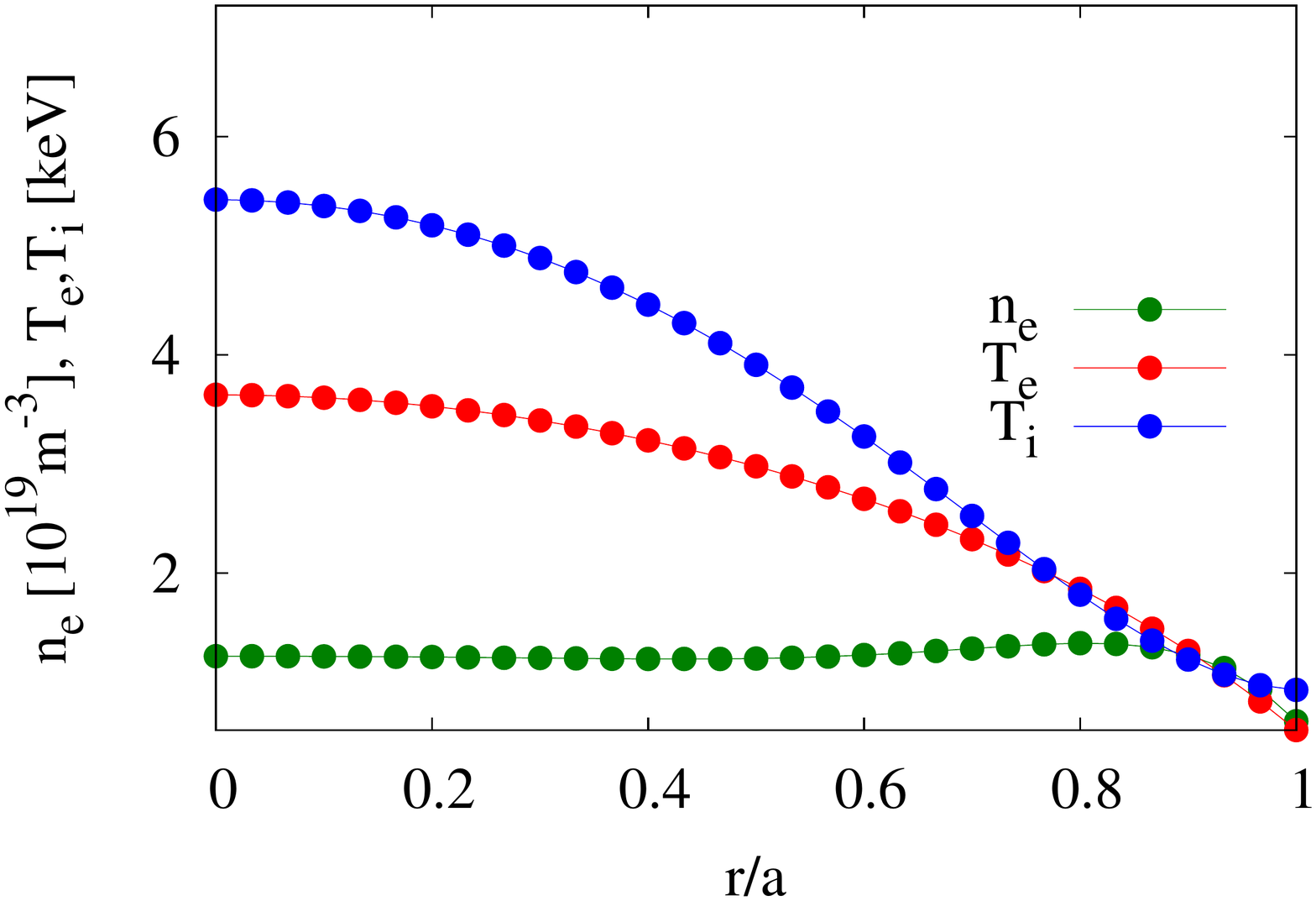}\vskip-2.5cm
\includegraphics[width=\columnwidth,angle=0]{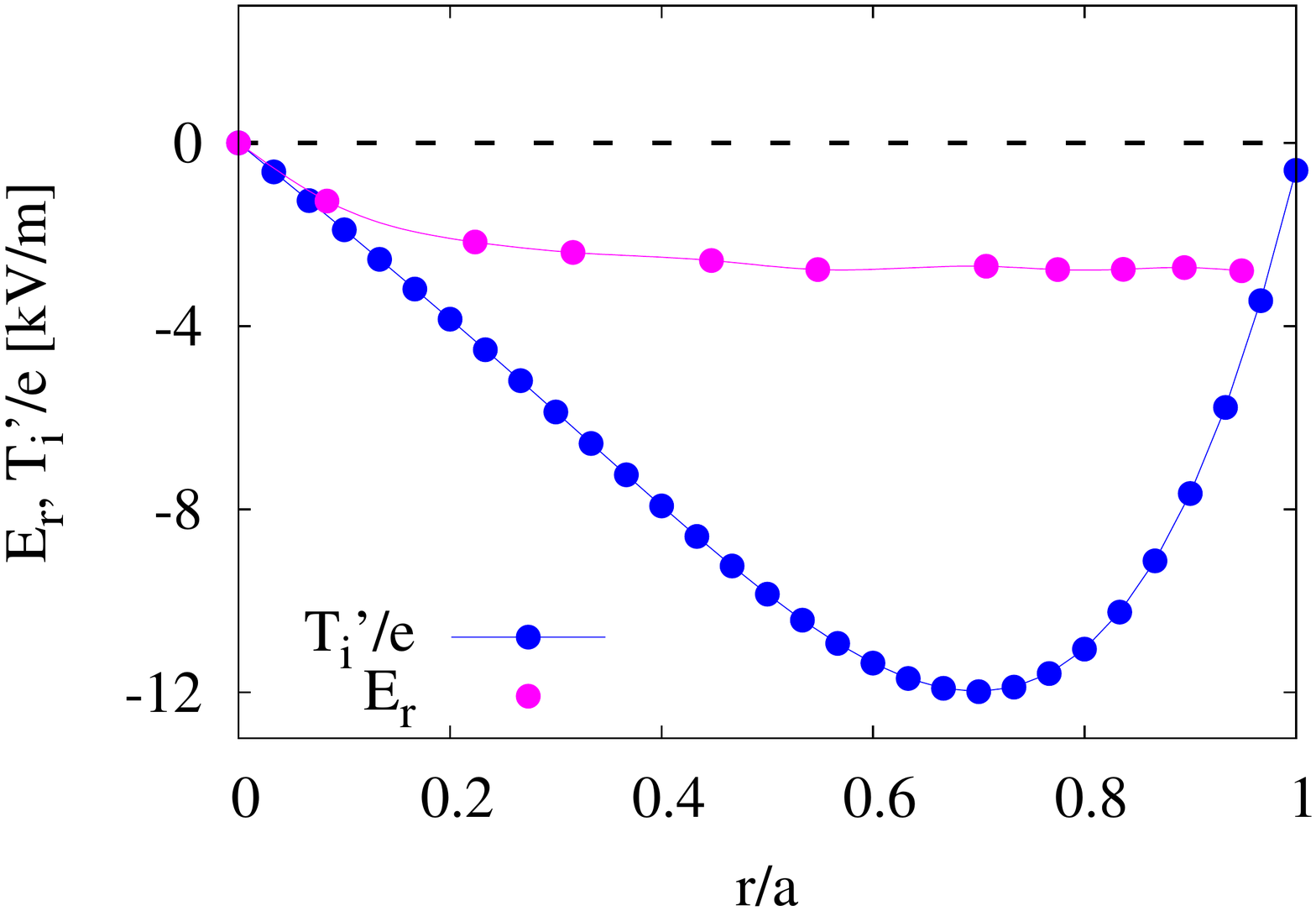}
\end{center}
\vskip-1cm\caption{Plasma profiles for a impurity hole plasma of LHD (top) and calculated radial electric field (bottom).}
\label{FIG_IH}
\end{figure}

\subsection{Examples}\label{SEC_EXAMPLES}

Let us illustrate the discussion of the previous section with two examples, represented in figures~\ref{FIG_IR} and \ref{FIG_IH}. The first one, shown in figure~\ref{FIG_IR} (top) is a typical ion root plasma taken from the International Stellarator/Heliotron Database~\footnote{\url{https://ishpdb.ipp-hgw.mpg.de/}\\\url{http://ishpdb.nifs.ac.jp/index.html}}, see also~\cite{dinklage2013ncval}. It is a plasma of high injected input power at medium densities, $6\times 10^{19}$m$^{-3}$ at the core, and both the ion and electron temperatures are around 2$\,$keV. The second example, shown in figure~\ref{FIG_IH} (top), is an impurity hole plasma~\cite{nunami2016iaea}: the density at the core is much lower, $1\times 10^{19}$m$^{-3}$ and the temperatures higher, close to 5$\,$keV. Since the shape of the profiles is similar in the two cases (hollow density, peaked temperatures), the main difference is the factor 20 in the collisionality of the core region. And the consequence, in line with what we have discussed in the previous section, is that the values of \jlvg{$eE_r$, when compared with the corresponding ion temperature gradient, are very different. In the ion root plasma, figure~\ref{FIG_IR} (bottom), $eE_r$ is basically equal to the ion temperature gradient; in the impurity hole plasma, figure~\ref{FIG_IH} (bottom), it is negative but much smaller in size than the ion temperature gradient. }

\begin{figure}
\begin{center}
\includegraphics[width=\columnwidth,angle=0]{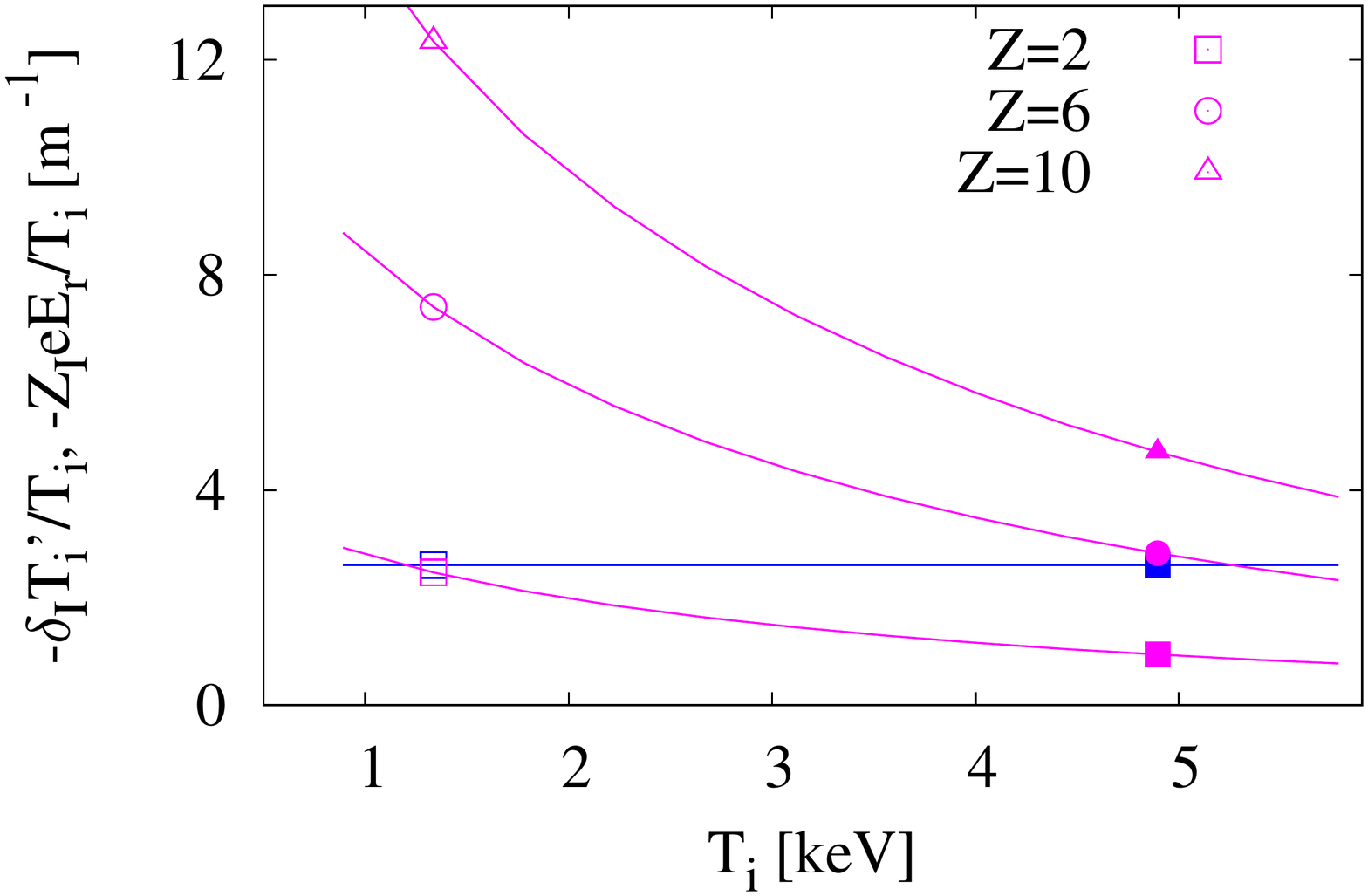}
\end{center}
\vskip-1cm\caption{Convective terms in the impurity flux as a function of the ion temperature. Open and closed signs correspond to the ion root and impurity hole plasma respectively. Here, $\delta_I\equiv L_2^I/L_1^I$.}
\label{FIG_SCAN}
\end{figure}

The effect on radial impurity transport can be seen in a collisionality scan depicted in figure~\ref{FIG_SCAN}, where the electron and ion temperatures are increased while keeping the density and all of the gradients constant: the driving term associated to the radial electric field ($Z_IE_r/T_i$, with $E_r$ calculated with \texttt{DKES}) is compared with the normalized ion temperature gradient for several charge-states. In the left part of the figure the plasma is in ion root and, of the two terms, the one containing the radial electric field clearly dominates impurity transport; as one moves to the right, to higher temperatures, the two terms start to compete. At the highest temperature, there would be an outwards pinch for $Z_I=2$, almost no pinch for $Z_I=6$, and inwards but small pinch for $Z_I=10$. We note that the specific details of figure~\ref{FIG_SCAN} depend among other things upon the choices of profiles; but the robust result is that at high temperatures, there is impurity screening for some impurities, and that the higher the temperature, the closer to temperature screening. In these situations, an additional contribution to the radial flux may determine whether impurities accumulate or they are expelled. In the next section, we discuss one of the possible physical mechanisms.

\section{\ivan{The effect of the tangential magnetic drift on the calculation of the} tangential electric field}\label{SEC_LARGEEPAR}

The variation of the electrostatic potential on the flux surfaces,
\begin{equation}
\varphi_1 = \varphi - \varphi_0(r),
\end{equation}
and its impact on the impurity flux has received much attention lately. The recent work~\cite{regana2017phi1} summarizes most of what has been learnt by means of numerical simulations (experimental validation has been attempted in~\cite{pedrosa2015phi1,liu2017phi1}) with the codes \texttt{EUTERPE}~\cite{regana2013euterpe} and \texttt{SFINCS}~\cite{landreman2014sfincs}: the electrostatic potential was calculated for a series of devices (LHD~\cite{ida2015lhd}, W7-X~\cite{sunnpedersen2016nature,klinger2017op11} and TJ-II~\cite{sanchez2015iaeap}) and plasma profiles. The conclusion was that taking into account this effect changes quantitatively the impurity flux, but it remains negative, inward-directed, in most of the cases studied. Furthermore, the size of the tangential electric field was shown to depend on the degree of optimization of the magnetic configuration, being smaller for stellarators with smaller ripple, which casted doubts on the relevance of this effect for a hypothetical stellarator reactor. 

In subsection \ref{SEC_EQUATIONS2} we recall how the tangential electric field is included in \texttt{EUTERPE} and \texttt{SFINCS} and why, if one is interested in dealing with situations in which the radial electric field is small, one also has to include the effect of the tangential magnetic field for the correct calculation of the tangential electric field. Then, we briefly describe the equations derived in \cite{calvo2017sqrtnu}, valid for stellarators close to omnigeneity when the non-omnigeneous perturbation has small gradients, that rigorously include the tangential electric field and the tangential magnetic drift. In subsection (\ref{SEC_KNOSOS}), calculations employing the code \texttt{KNOSOS}, that solves the equations of reference \cite{calvo2017sqrtnu}, are given. The discussion in this section is carried out assuming that the electrons are adiabatic, as in previous works~\cite{regana2017phi1}, which will allow us to single out the effect of the tangential magnetic drift on the calculation of $\varphi_1$. Simulations including kinetic electrons in the quasineutrallity equation, and not only in the ambipolarity equation, will be presented elsewhere.

\subsection{Equations}\label{SEC_EQUATIONS2}

The codes \texttt{EUTERPE} and \texttt{SFINCS} solve (e.g. in~\cite{regana2017phi1}) a drift-kinetic equation that, at low collisionality, is equivalent to
\begin{equation}
\overline{\mathbf{\ivan{v_{E,0}}}\cdot\nabla\alpha}~\partial_\alpha g_b +\overline{(\mathbf{\mathbf{v_M}}+\mathbf{v_E})\cdot\nabla r}~F_{M_b} \Upsilon_b = \overline{C(g_b)}\,.\label{EQ_DKE2}
\end{equation}
The difference between this equation and equation (\ref{EQ_DKE}) is that in (\ref{EQ_DKE2}) we have included the orbit-average of the radial component of the $\mathbf{E}\times \mathbf{B}$ drift. This drift is caused by the tangential electric field associated to the variation of the electrostatic potential on the flux surface $\varphi_1$. The latter is calculated by solving the quasineutrality equation\jlvg{; retaining only the adiabatic response of the electrons (as in equation (20) of reference~\cite{regana2017phi1}), it} reads
\begin{equation}
\left(\frac{Z_i}{T_i}+\frac{1}{T_e}\right)\varphi_1 =\frac{2\pi}{en_e} \int_0^\infty\mathrm{d}v \int_{B^{-1}_{{\mathrm{max}}}}^{B^{-1}}\mathrm{d}\lambda\frac{v^3 B}{|v_{||}|} g_i\,.\label{EQ_QN}
\end{equation}
We note that equation~(\ref{EQ_DKE2}) is able to describe only the $1/\nu$ or $\sqrt{\nu}$ regimes of neoclassical transport of the bulk ions. However, at low collisionalities, and especially for small radial electric fields, the contribution from the so-called superbanana-plateau regime to the transport of bulk ions cannot be neglected. This has been long known for radial transport~\cite{matsuoka2015tangential,calvo2017sqrtnu} and for the parallel flows~\cite{huang2017bootstrap}; recently~\cite{calvo2017sqrtnu}, it has been predicted to have a strong impact on the electrostatic potential variations on the flux-surface. We use some of these results in what follows. 

The superbanana-plateau regime is not captured by (\ref{EQ_DKE2}) (and neither by equation (\ref{EQ_DKE}), of course) because the first term on the left side contains the orbit-average of the tangential component of the $\mathbf{E}\times \mathbf{B}$ drift but not of the tangential magnetic drift. If the aspect ratio is large and the radial electric field is not too small, the tangential magnetic drift is negligible with respect to the tangential $\mathbf{E}\times \mathbf{B}$ drift. However, at small collisionality, it cannot be neglected when the radial electric field is small, even if the stellarator has large aspect ratio. \ivan{In~\cite{calvo2017sqrtnu}} it has been explained that, in this situation, the neoclassical equations remain radially local and linear only if \ivan{the magnetic configuration is sufficiently optimized, meaning that the orbit averaged radial magnetic drift is small for all trapped trajectories. In other words, if the stellarator is sufficiently close to omnigeneity}.

We say that the magnetic configuration is close to omnigeneity if $B$ can we written as
\begin{equation}
B=B_0+ B_1,
\end{equation}
where $B_0$ is omnigenous and $B_1$ is a small non-omnigeneous \ivan{perturbation}. In reference \cite{calvo2017sqrtnu}, linear, radially local  drift-kinetic and quasineutrality equations have been derived~\jlvg{when $B_1$ is small and} has small gradients. The dominant contributions to $g_i$ and $\varphi_1$ come from the non-omnigeneous perturbation $B_1$ and can be determined by solving the drift-kinetic equation
\begin{equation}\label{eq:DKE_KNOSOS}
\overline{(\mathbf{v_M} + \mathbf{v_E})\cdot\nabla\alpha}^{(0)}~\partial_\alpha g_i +\overline{(\mathbf{\mathbf{v_M}}+\mathbf{v_E})\cdot\nabla r}^{(1)}~F_{M_i} \Upsilon_i = \overline{C(g_i)}^{(0)}\,,
\end{equation}
and the quasineutrality equation
 \begin{equation}\label{eq:QN_KNOSOS}
\left(\frac{Z_i}{T_i}+\frac{1}{T_e}\right)\varphi_1 =\frac{2\pi}{en_e} \int_0^\infty\mathrm{d}v \int_{B^{-1}_{{0,\mathrm{max}}}}^{B_0^{-1}}\mathrm{d}\lambda\frac{v^3 B_0}{|v_{||}^{(0)}|} g_i\,.
\end{equation} 
Here, $B_{0,\mathrm{max}}$ is the maximum value of $B_0$ on the flux surface,
\begin{equation}
| v_\parallel^{(0)} |(r,\alpha,l,v,\lambda) = v\sqrt{1-\lambda B_0(r,\alpha,l)} \, , 
\end{equation}
\begin{equation}\label{eq:average_tangential_drift}
\fl  \hspace{1cm}\overline{(\mathbf{v_M} + \mathbf{v_E})\cdot\nabla\alpha}^{(0)}
  =
  \frac{m_i}{Z_i e \Psi'_t \tau^{(0)}}
  \int_{l_{b_{10}}}^{l_{b_{20}}}
\frac{\lambda v\partial_r B_0 + 2Z_i e/(m_iv)\varphi'_0}
{\sqrt{1-\lambda B_0}}\mathrm{d} l,
\end{equation}
\begin{equation}\label{eq:average_tangential_drift2}
\fl \hspace{1cm}\overline{(\mathbf{\mathbf{v_M}}+\mathbf{v_E})\cdot\nabla r}^{(1)}=
-\frac{m_i}{Z_i e \Psi'_t \tau^{(0)}}
\partial_\alpha
\int_{l_{b_{10}}}^{l_{b_{20}}}
\frac{
\lambda v B_1 + 2Z_i e/(m_iv)\varphi_1
}
{\sqrt{1-\lambda B_0}} \mathrm{d} l, 
\end{equation}
$l_{b_{10}}$, $l_{b_{20}}$ are the bounce points of the trapped orbit in the magnetic field $B_0$, determined by the equation $1-\lambda B_0 = 0$,
\begin{equation}
\tau^{(0)} = 2\int_{l_{b_{10}}}^{l_{b_{20}}}|v_\parallel^{(0)}|^{-1}\mathrm{d} l
\end{equation}
is the time that it takes for a particle to complete the orbit in the magnetic field $B_0$ and
\begin{equation}
\overline{C(g_i)}^{(0)} = \frac{2}{\tau^{(0)}}\int_{l_{b_{10}}}^{l_{b_{20}}}|v_\parallel^{(0)}|^{-1} C^{(0)}(g_i)\mathrm{d} l,
\end{equation}
where $C^{(0)}(g_i)$ is the pitch-angle-scattering operator corresponding to $B_0$. We have ended up with a a system of equations that is radially local, that includes the tangential magnetic drift, and that is linear in the unknowns, $g_i$ and $\varphi_1$. 

\begin{figure}
\begin{center}
\includegraphics[width=0.41\columnwidth,angle=270]{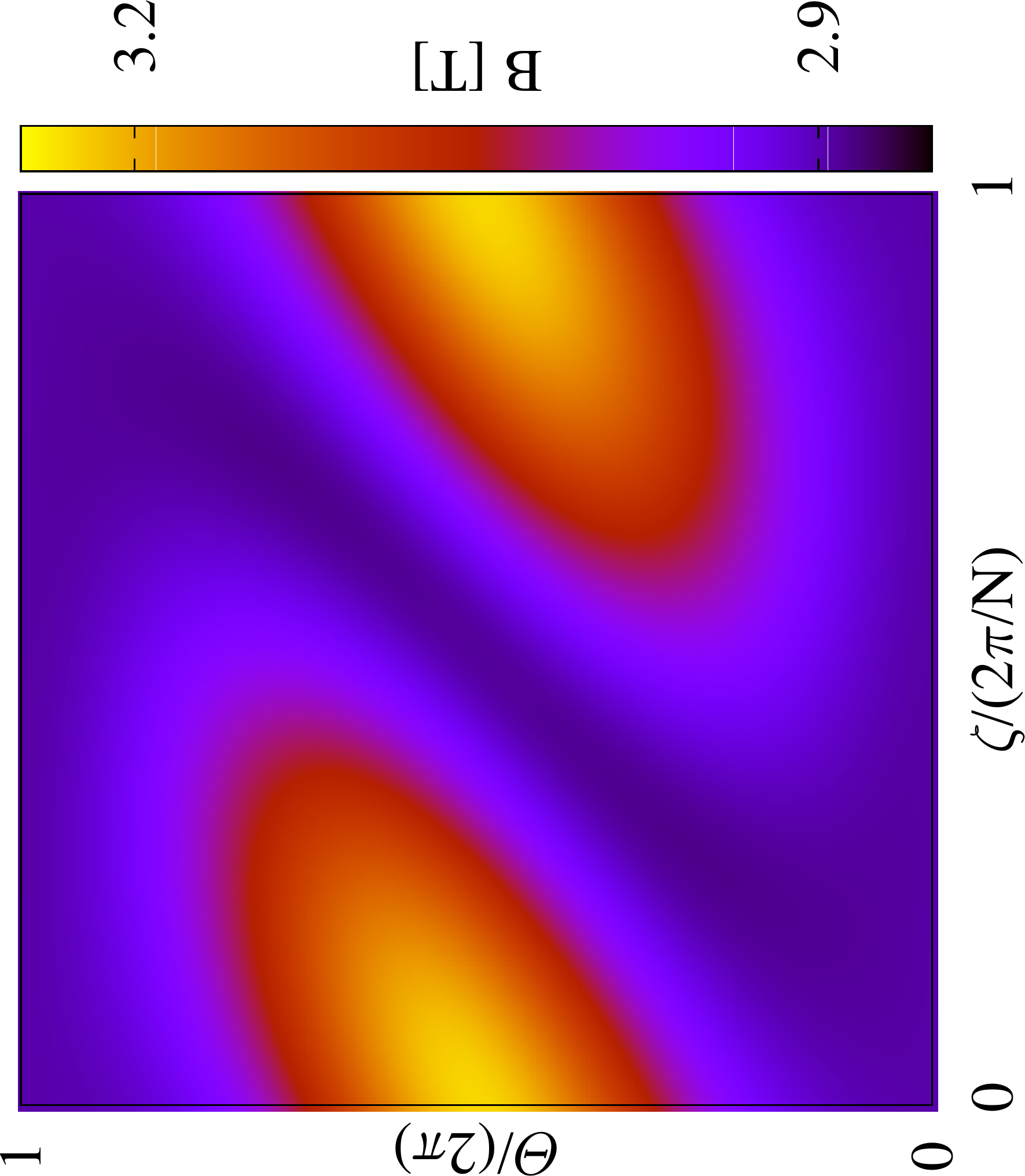}
\includegraphics[width=0.41\columnwidth,angle=270]{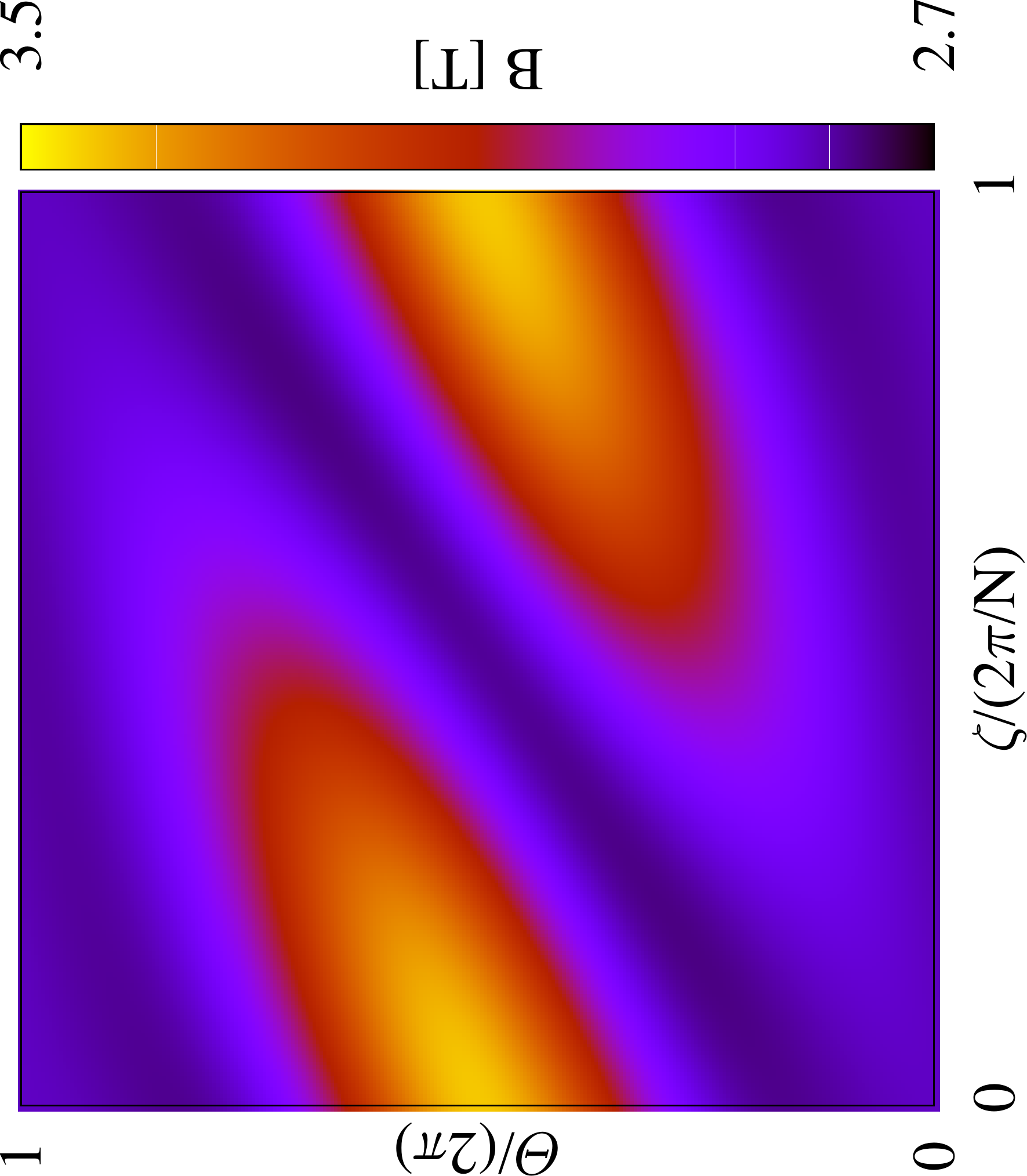}

\

\includegraphics[width=0.41\columnwidth,angle=270]{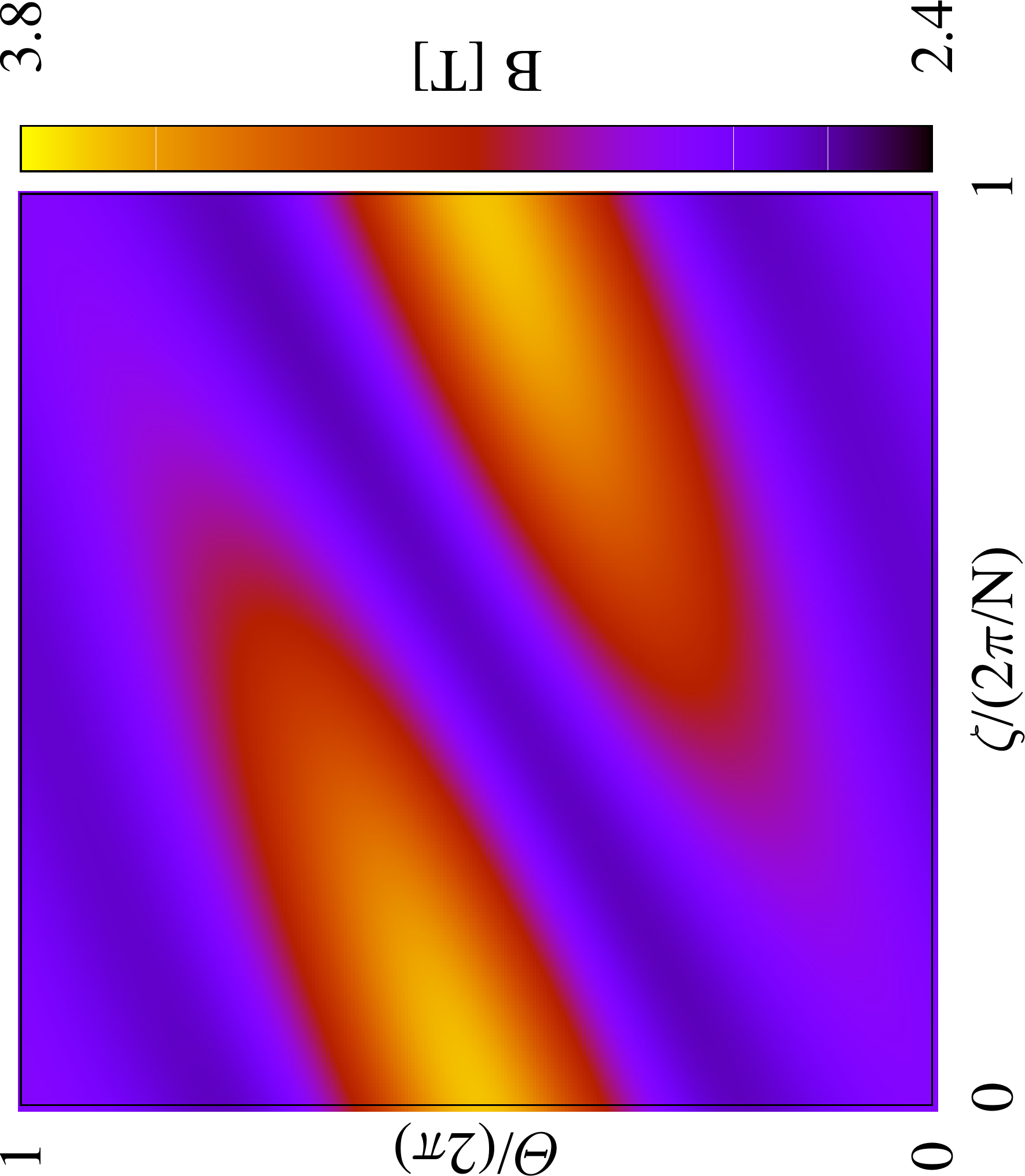}
\includegraphics[width=0.41\columnwidth,angle=270]{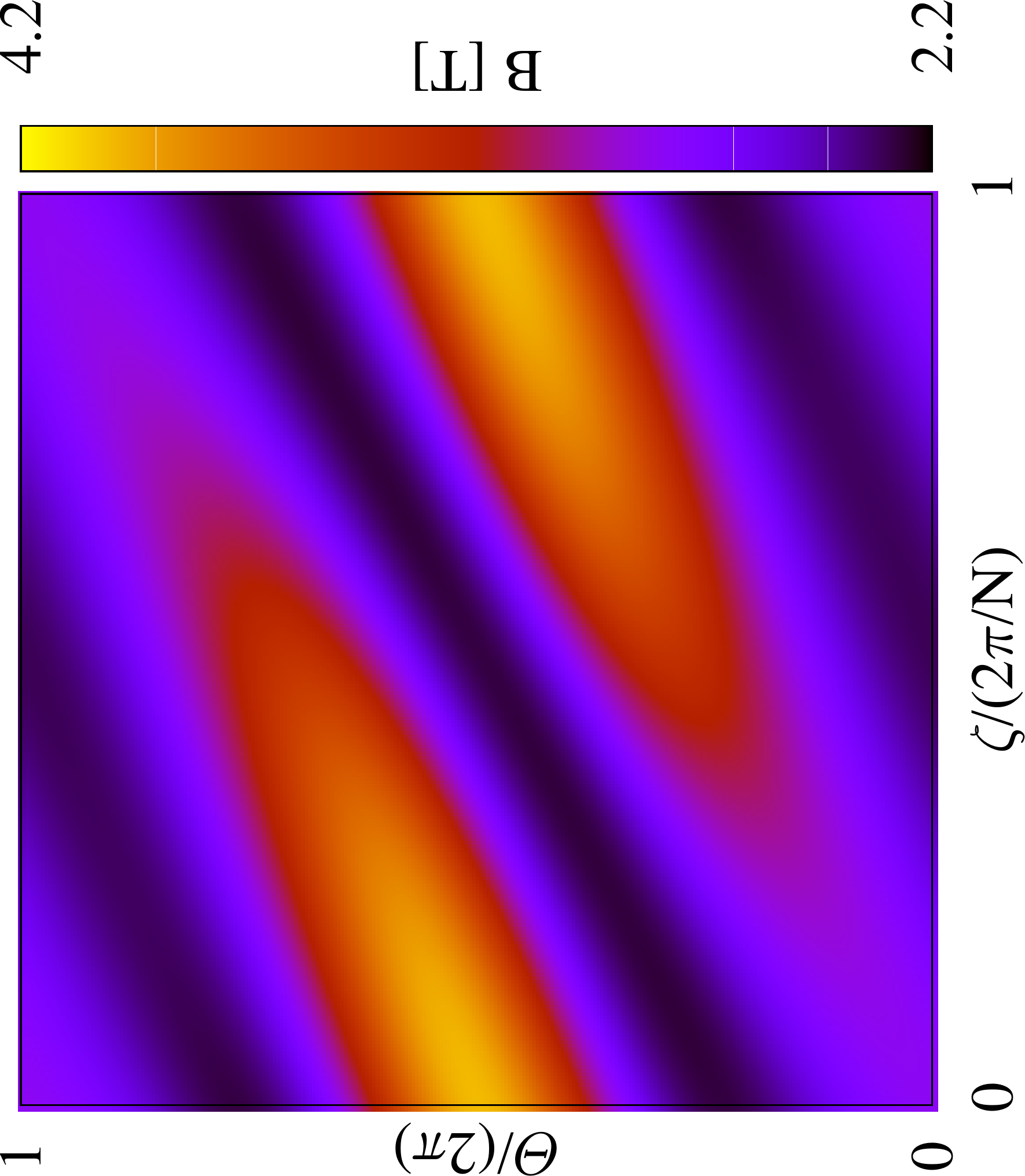}
\end{center}
\caption{Magnetic field strength as a function of the toroidal and poloidal Boozer angles, $\zeta$ and $\Theta$, for $r/a\,=\,$0.2 (top left), 0.4 (top right), 0.6 (bottom left) and 0.8 (bottom right).}
\label{FIG_BFIELD}
\end{figure}

\begin{table}
\begin{center}
\begin{tabular}{c||ccc|ccc|ccc|ccc}
$r/a$ & & 0.2 & & & 0.4 & & & 0.6  & & & 0.8\\\hline\hline
 & & & & & & & & &  & \\
$\frac{|B_{m,n}|}{|B_{0,0}|}\!>\!10^{-2}$  & $m$ & $n$ & $B_{m,n}\,$[T] & $m$ & $n$ & $B_{m,n}\,$[T]  & $m$ & $n$ & $B_{m,n}\,$[T]  & $m$ & $n$ & $B_{m,n}\,$[T]\\
& & & & & & & & &  & \\\hline
& & & & & & & & &  & \\
	 	                                         &  0 & 0 & 2.975      & 0 & 0   & 2.945    & 0 & 0  & 2.902   & 0 &  0 & 2.826\\
$B_0$      				                & 1 &-1 &-0.08257  & 1 & -1 & -0.1394 & 2 & -1 & 0.2822 & 2 & -1 & 0.5196\\
& & & & & & & & &  & \\\hline
& & & & & & & & &  & \\
  						         & 1 & 0 &-0.08449  & 1 & 0  & -0.1617  & 1 &   0 & -0.2362   & 1 & 0   & -0.3114\\
$B_1$                                                & 0 & 1 & 0.04665  & 2 & -1 & 0.1346   & 1 & -1  & -0.1741   & 1 & -1  & -0.1943\\
 							& 2 & -1 & 0.03761 & 0 & 1  & 0.03320 & 3 & -1  & -0.05654 & 3 & -1 &-0.1190 \\
                                                           &    &     &                 &    &     &                 & 2 & 0   & 0.0299   & 2 & 0  & 0.04297\\
& & & & & & & & &  & \\\hline
\end{tabular}
\end{center}
\caption{Main harmonics of the Boozer representation of the magnetic field.}
\label{TAB_CONF}
\end{table}

\begin{figure}
\begin{center}
\includegraphics[width=0.41\columnwidth,angle=270]{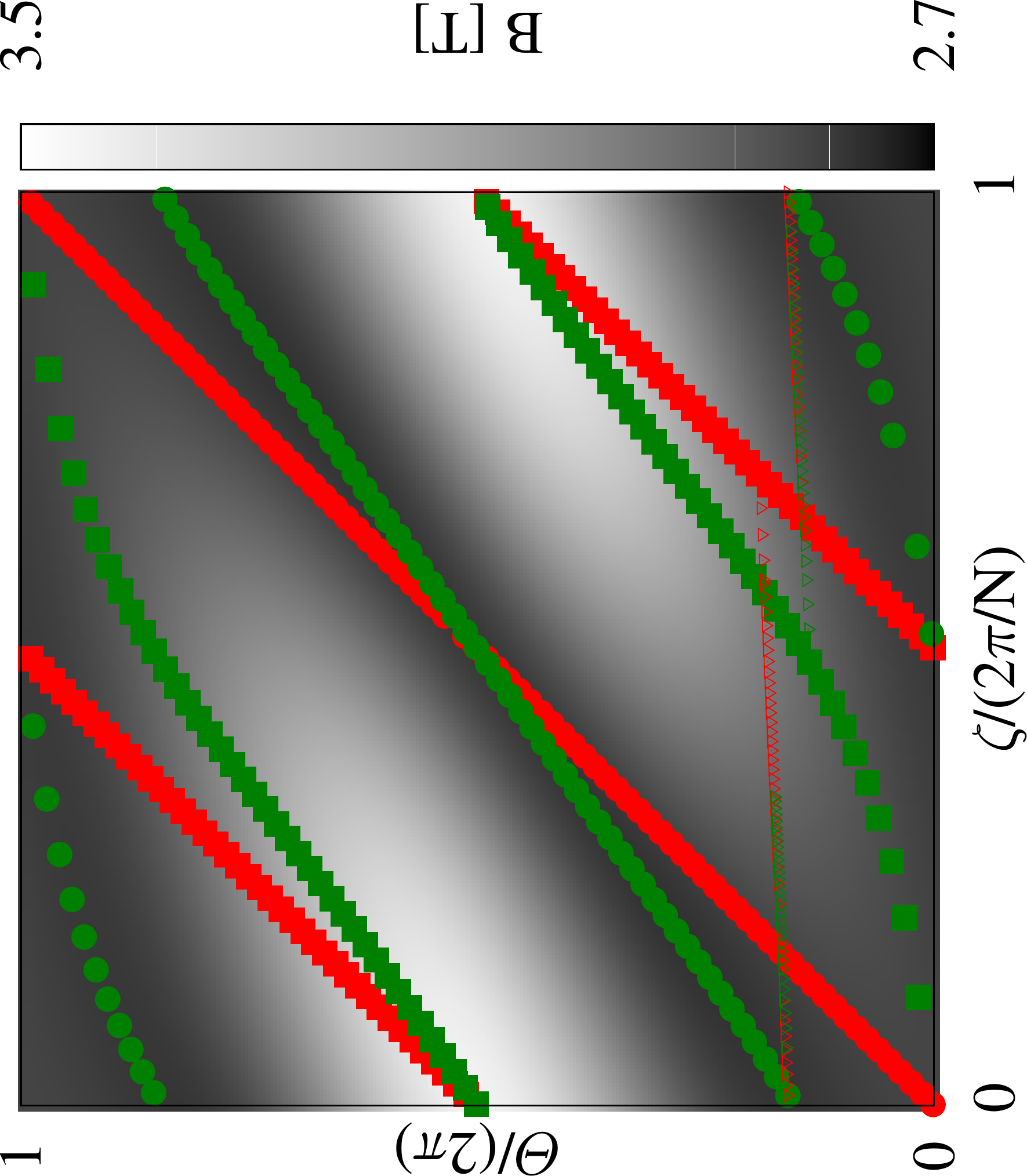}
\includegraphics[width=0.41\columnwidth,angle=270]{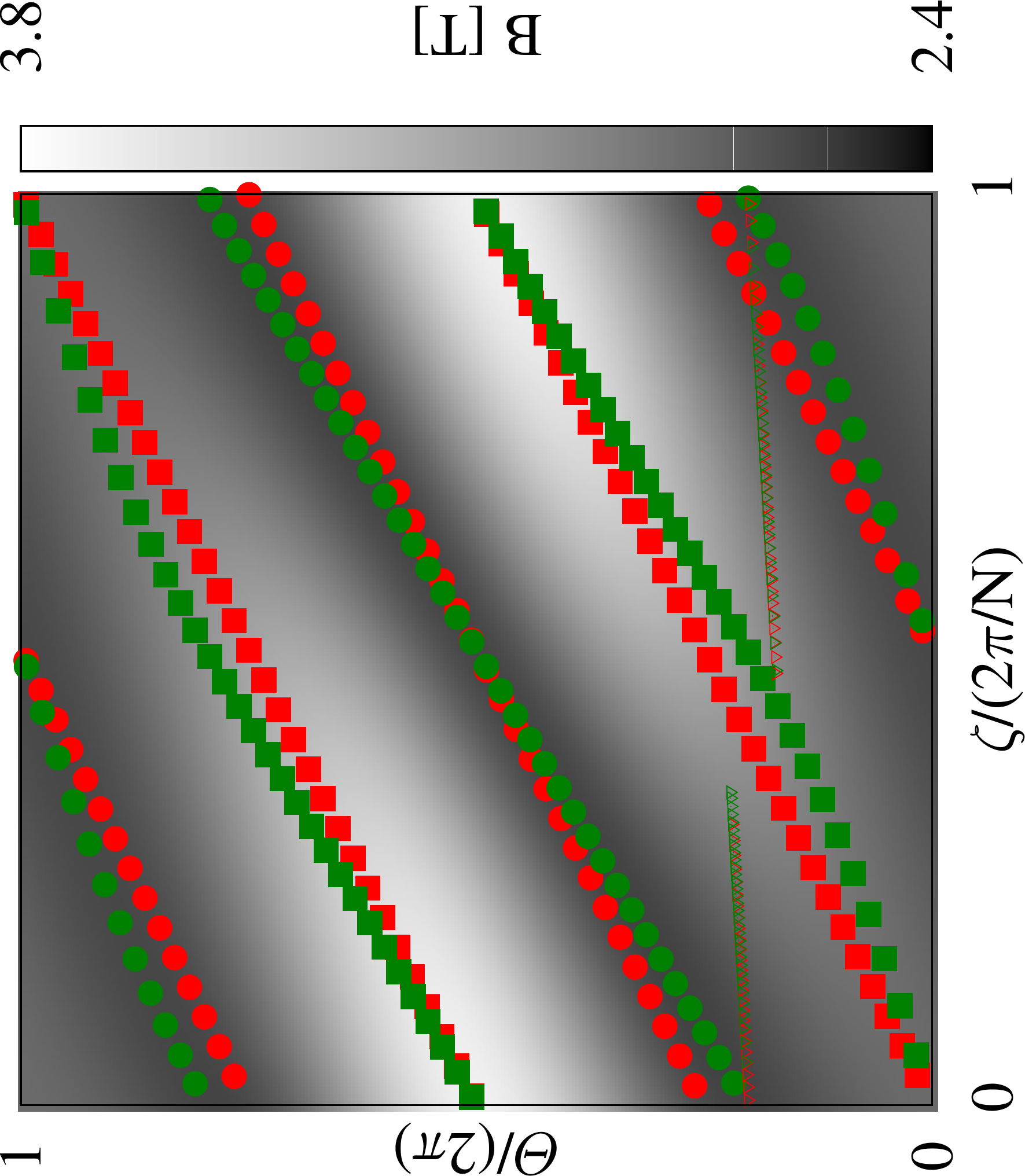}
\end{center}
\caption{Contours of maximum (squares) and minimum (circles) values of $B$ along the magnetic field line for $B_0$ (red) and $B_0+B_1$ (green) at $r/a=$0.2 (left) and 0.6 (right); the open triangles correspond to a selected field line.}
\label{FIG_BLINES}
\end{figure}

\begin{figure}
\begin{center}
\includegraphics[width=\columnwidth,angle=0]{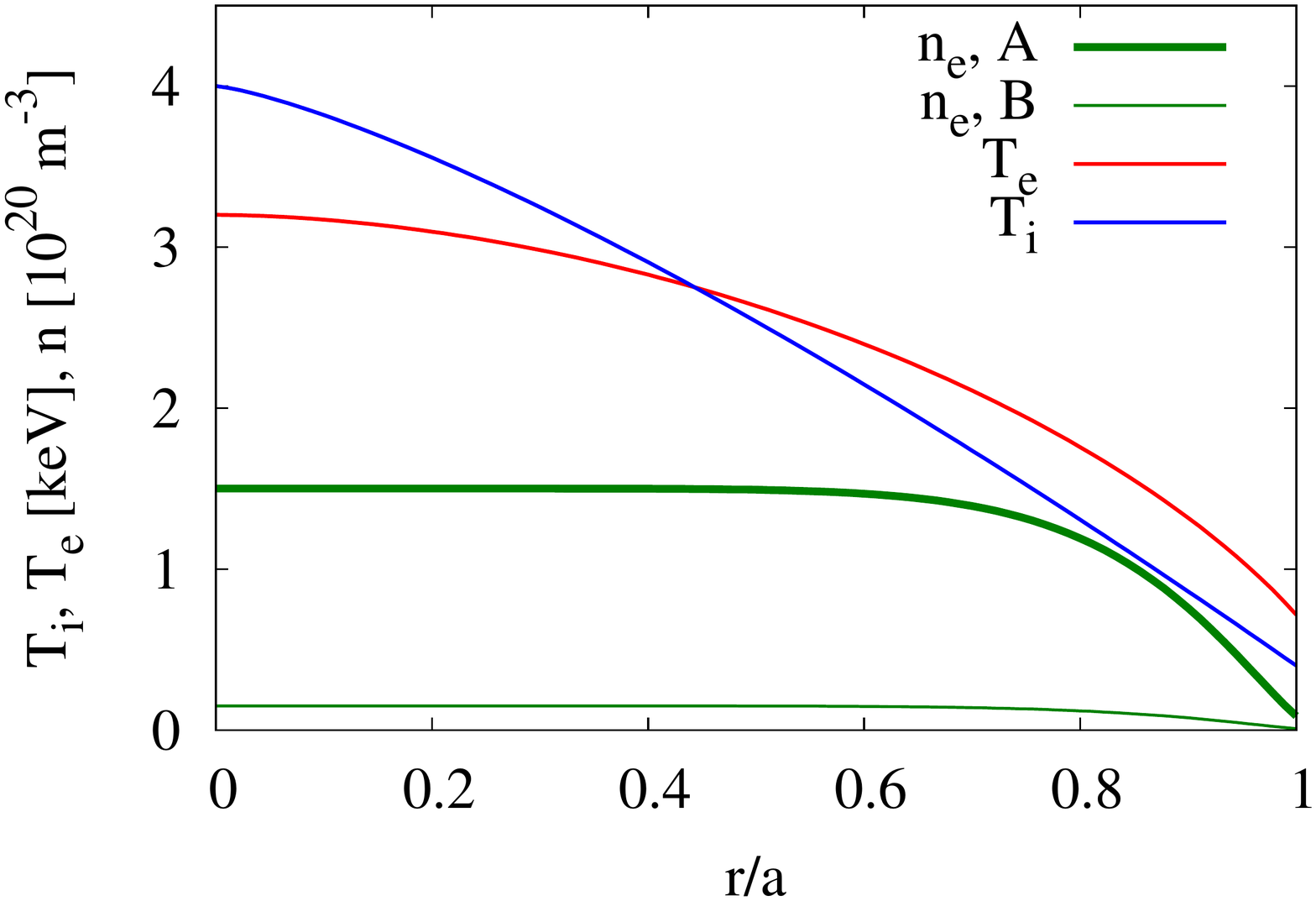}\vskip-2.5cm
\includegraphics[width=\columnwidth,angle=0]{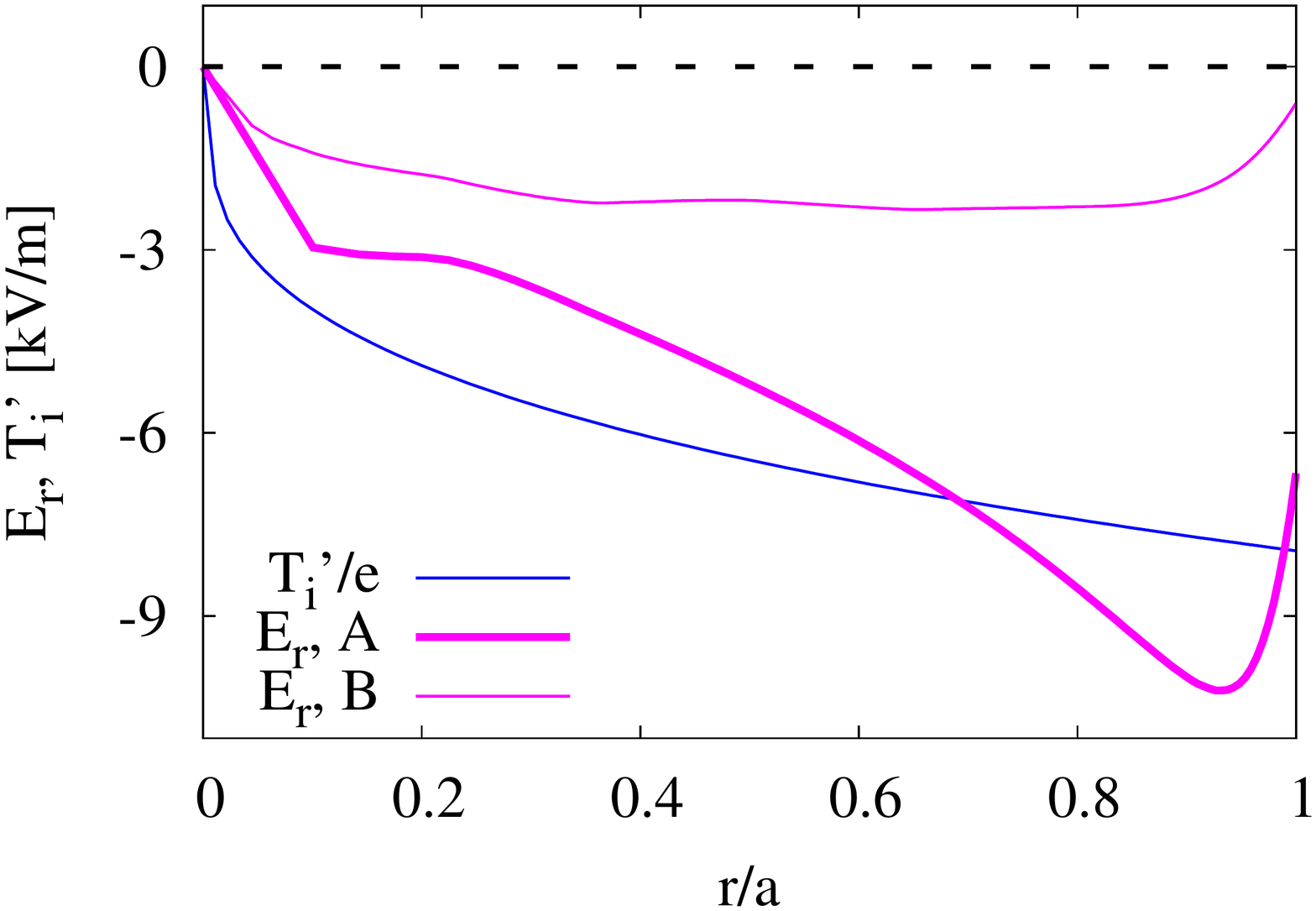}
\end{center}
\caption{Plasma profiles used for the calculation of $\varphi_1$.}
\label{FIG_PROFILES_PHI1}
\end{figure}

\subsection{\texttt{KNOSOS}}\label{SEC_KNOSOS}

We have written a code, the KiNetic Orbit-averaging SOlver for Sellarators (\texttt{KNOSOS}), that solves equations (\ref{eq:DKE_KNOSOS}) and (\ref{eq:QN_KNOSOS}), and we show in this paper first simulations of the electrostatic potential at low collisionalities including the superbanana-plateau contribution. Details of the code will be given elsewhere, and in this section we briefly discuss how the calculations are done with \texttt{KNOSOS} for the plasmas relevant for this paper. These are LHD plasmas corresponding to the inward-shifted ($R=3.60\,$m) configuration. Figure~\ref{FIG_BFIELD} shows the magnetic field strength $B$ as a function of the toroidal and poloidal Boozer angles, $\zeta$ and $\Theta$ ($N=10$ is the number of periods), for several flux-surfaces, corresponding approximately to $r/a\,=\,$0.2, 0.4, 0.6 and 0.8. The harmonics of the Boozer representation of the magnetic field, $B = \sum_{m,n} B_{mn} \cos{(m\Theta + nN\zeta)}$, with significant magnitudes are listed in table~\ref{TAB_CONF} for the four flux-surfaces. One can see that, \ivan{as is typical} for heliotrons, a small number of harmonics is required for describing the magnetic field. For this reason, the so-called three-helicity model is enough to compute accurately the solution of equation~\ref{EQ_DKE}, as shown e.g. in~\cite{beidler2011ICNTS} and references therein.

With this in mind, in order to apply the model given by equations (\ref{eq:DKE_KNOSOS}) and (\ref{eq:QN_KNOSOS}), we have written the LHD magnetic field as a quasisymmetric magnetic field $B_0$ (quasisymmetry is a particular case of omnigeneity~\cite{cary1997omni}) plus a perturbation, and table~\ref{TAB_CONF} already shows how the harmonics have been grouped in $B_0$ and $B_1$ at each flux-surface: at $r/a\,<\,0.5$, we have chosen $B_0\!=\!B_0(\Theta-N\zeta)$, while for  $r/a\,>\,0.5$, we have set $B_0\!=\!B_0(2\Theta-N\zeta)$. \jlvg{We immediately note that, if we define our expansion parameter as $\delta \equiv B_1 / (B_0-B_{00})$, table~\ref{TAB_CONF} shows that $\delta\ll 1$ is not fulfilled, \ivan{especially for $r/a\,<\,0.5$, and therefore the modelling of $\varphi_1$ based on the $\delta$ expansion might be quantitatively inaccurate. Strictly speaking, when $\delta\sim 1$ and $E_r$ is small, one cannot justify a local theory of neoclassical transport at very low collisionality, and radially global neoclassical simulations may be required in order to calculate radial transport and $\varphi_1$~\cite{calvo2017sqrtnu}. Still, our local model represents a better approximation to reality than other local models that drop the tangential magnetic drift.}}

\jlvg{\ivan{We proceed} in two steps. We first switch-off the tangential magnetic drift in equation~(\ref{eq:DKE_KNOSOS}), yielding it equivalent to equation~(\ref{EQ_DKE2}): it can then solved without resorting to closeness to omnigeneity~\cite{hokulsrud1987sqrtnu}, and we use it for benchmarking \texttt{KNOSOS} against \texttt{EUTERPE}. Then we solve the complete equation~(\ref{eq:DKE_KNOSOS}) with \texttt{KNOSOS}: wherever \ivan{the solutions are significantly different than those of equation~(\ref{EQ_DKE2}), this will be an indication of the fact that the local approximation does not hold, and that radially global neoclassical simulations including the tangential magnetic drift might be needed for accurate results}. The \ivan{main conclusion} of this part of the paper will be to show that this is the situation for LHD experimentally-relevant plasmas.}

\jlvg{Finally, once we have noted that $\delta\sim 1$ at some radial positions, several choices of $B_0$ and $B_1$ are actually possible. The choices mentioned in the previous paragraphs are spurred by figure~\ref{FIG_BLINES}, where we plot the flux-surface maps of figure~\ref{FIG_BFIELD} emphasizing the details that are going to be relevant for the bounce-average calculations. For two radial positions, we plot the angular location of the maxima and minima of the magnetic field strength \textit{along several magnetic field lines}, and a selected field line, both for $B$ and for $B_0$. We see that the maxima of $B_0$ lie closer to those of $B$ than if we had chosen e.g. $B_0=B_0(\Theta)$.}

\begin{figure}
\begin{center}
\includegraphics[width=0.3\columnwidth,angle=270]{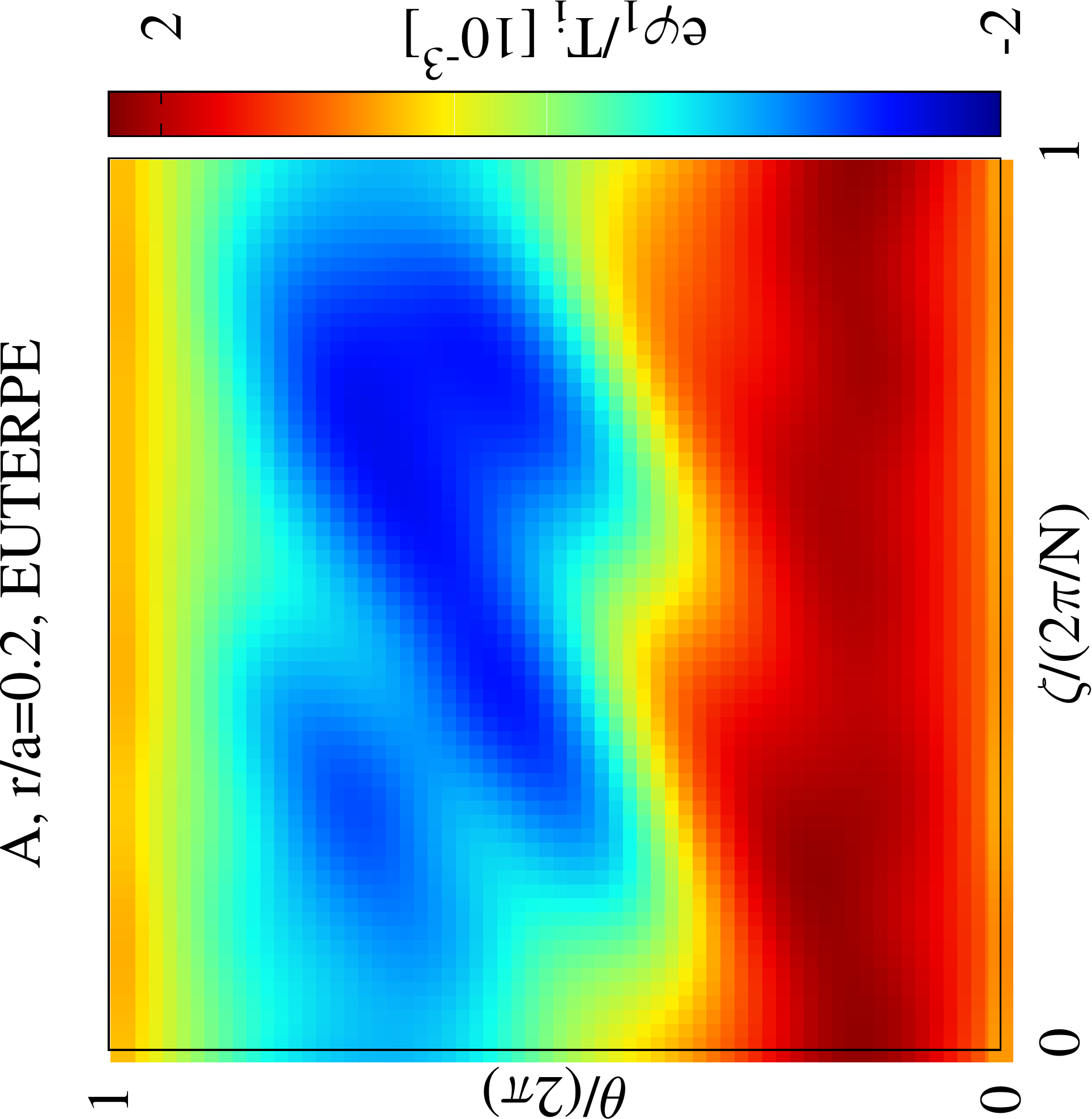}
\includegraphics[width=0.3\columnwidth,angle=270]{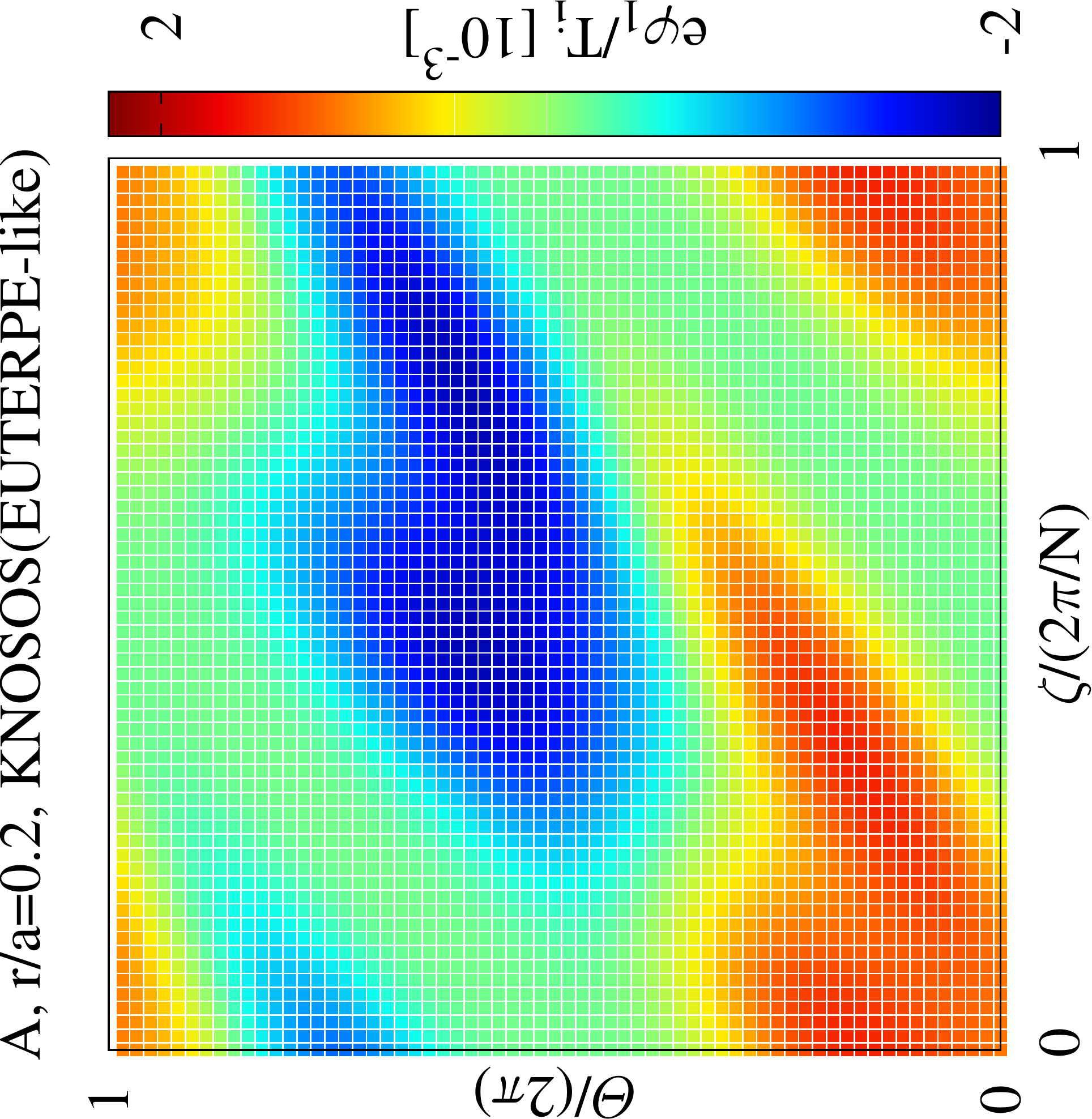}
\includegraphics[width=0.3\columnwidth,angle=270]{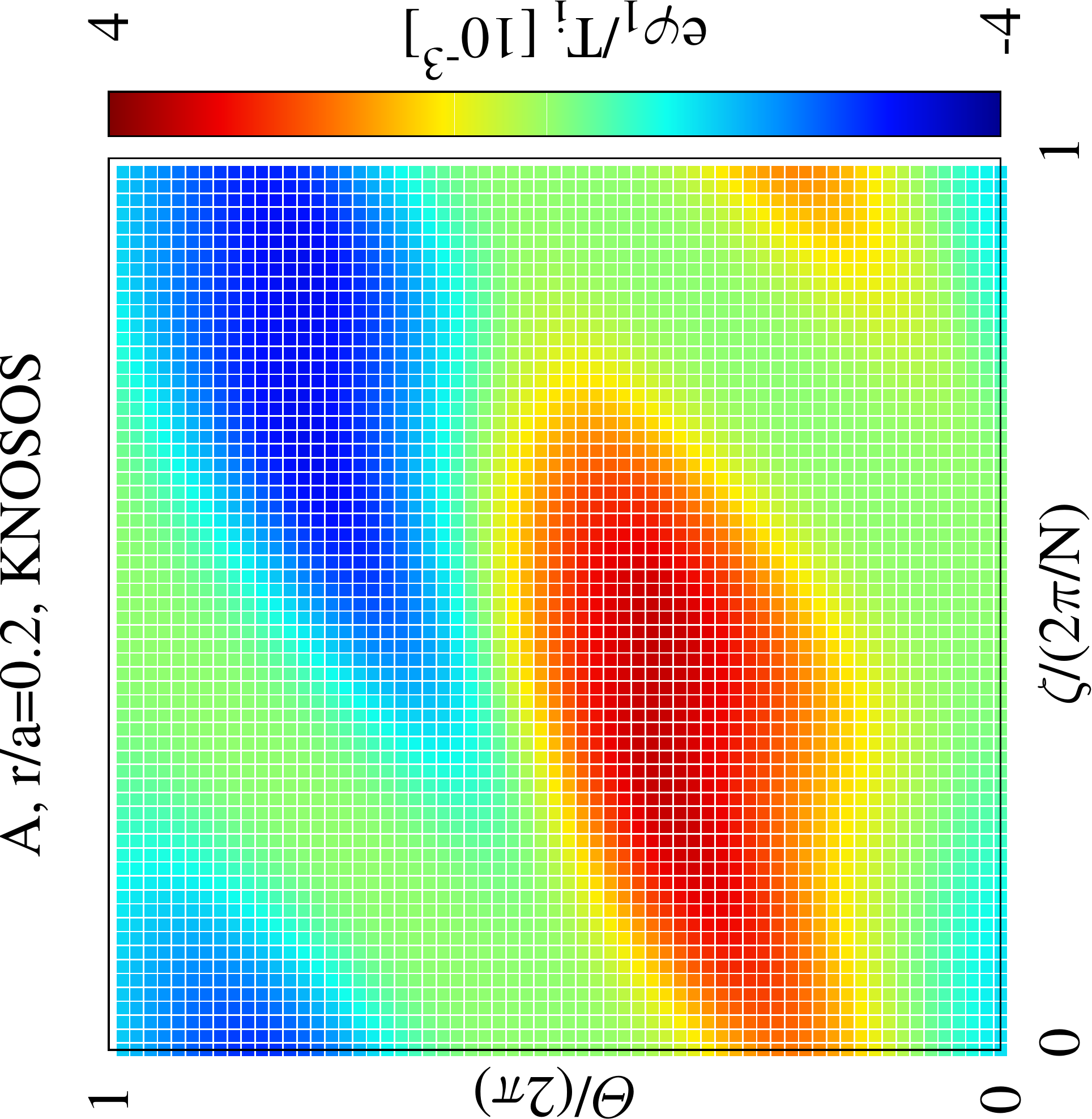}

\

\includegraphics[width=0.3\columnwidth,angle=270]{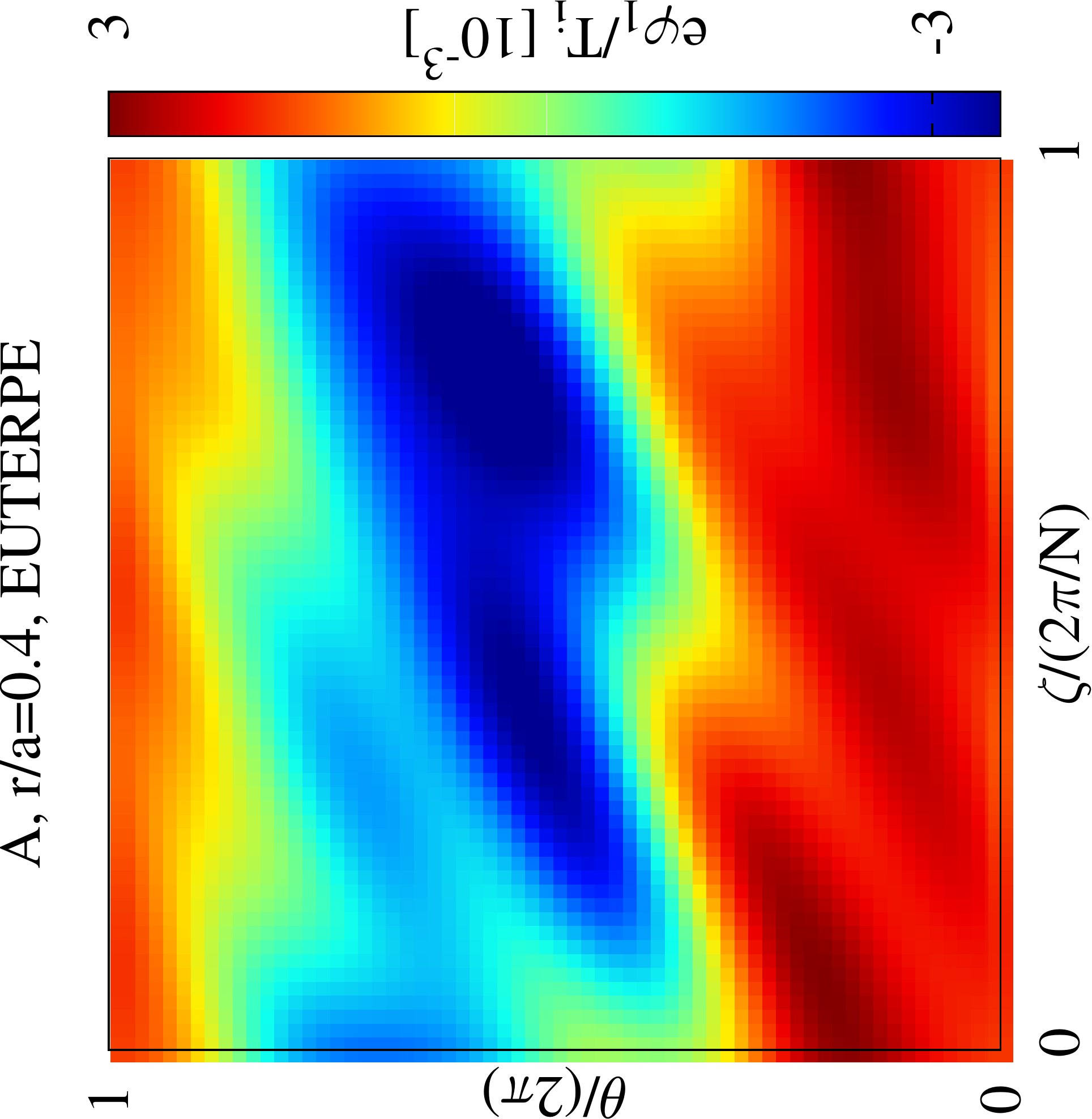}
\includegraphics[width=0.3\columnwidth,angle=270]{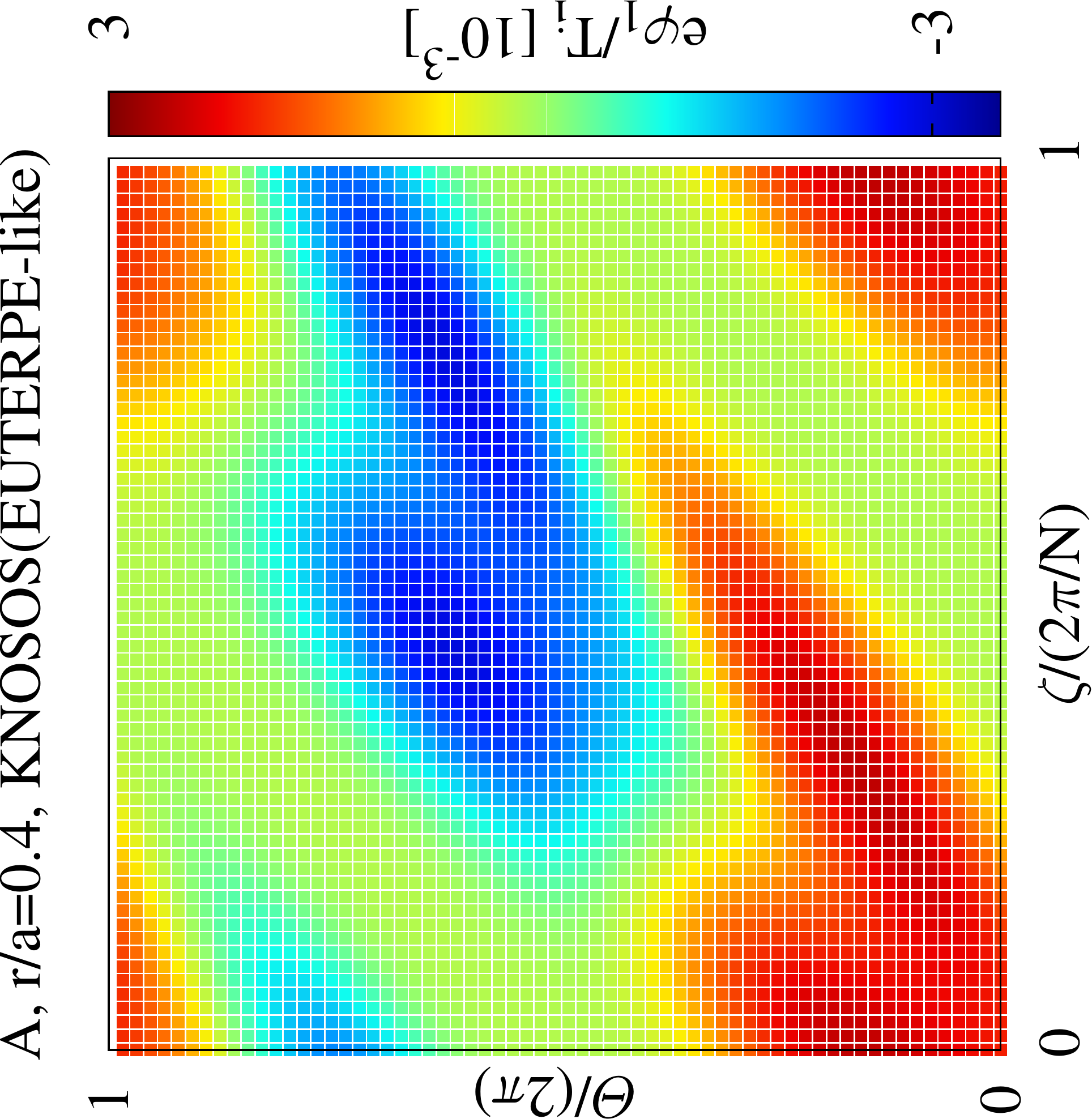}
\includegraphics[width=0.3\columnwidth,angle=270]{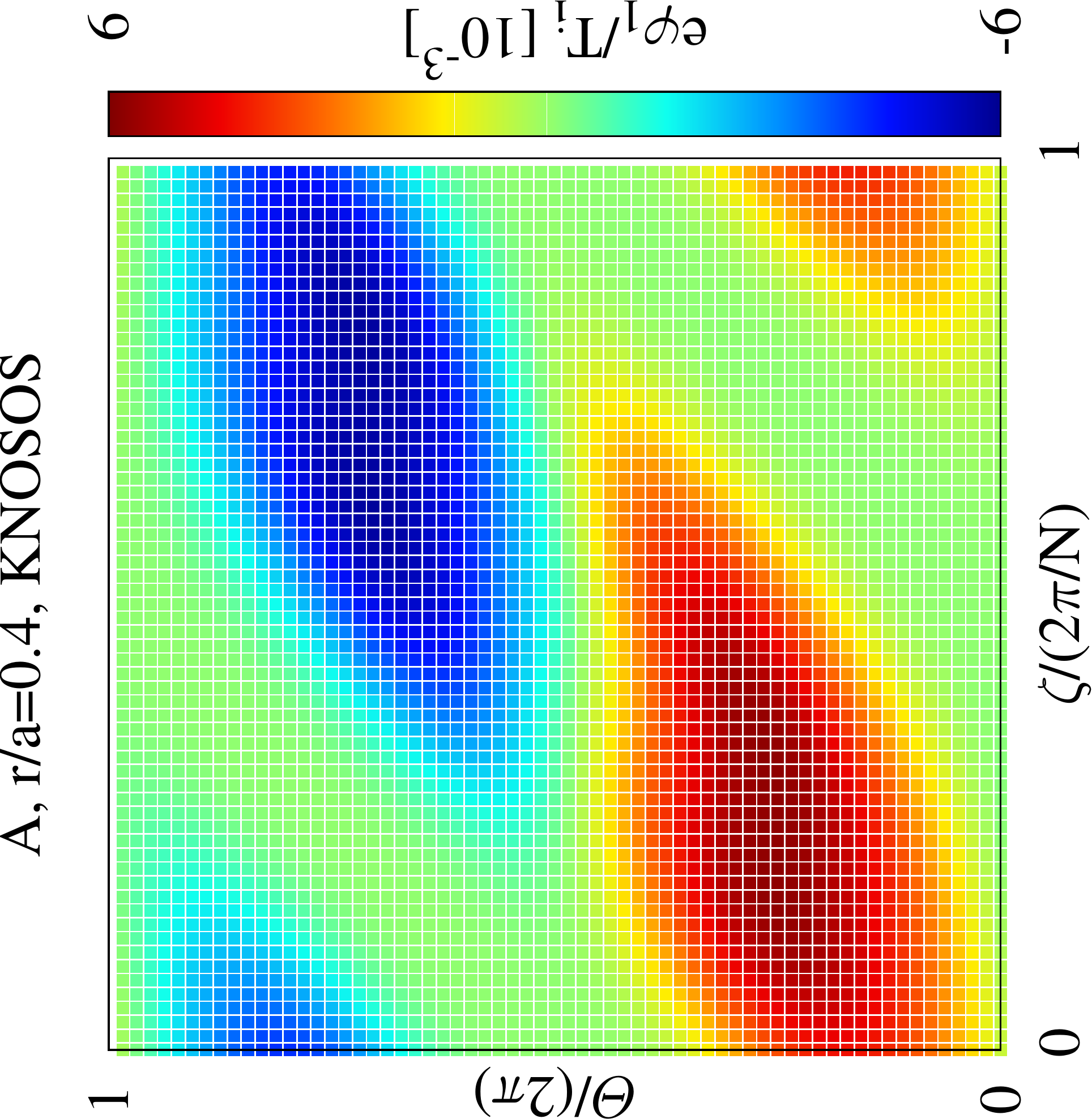}

\

\includegraphics[width=0.3\columnwidth,angle=270]{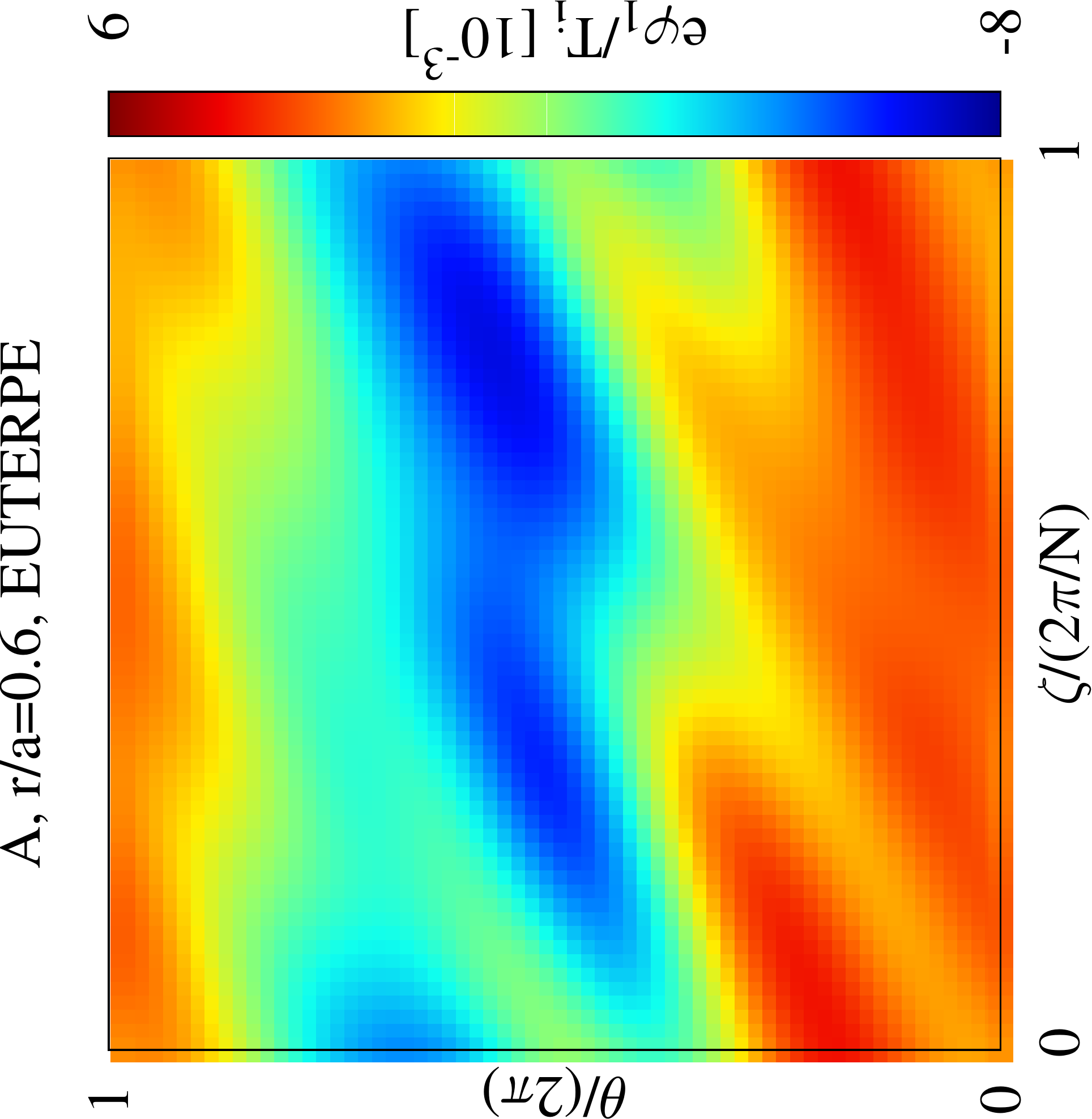}
\includegraphics[width=0.3\columnwidth,angle=270]{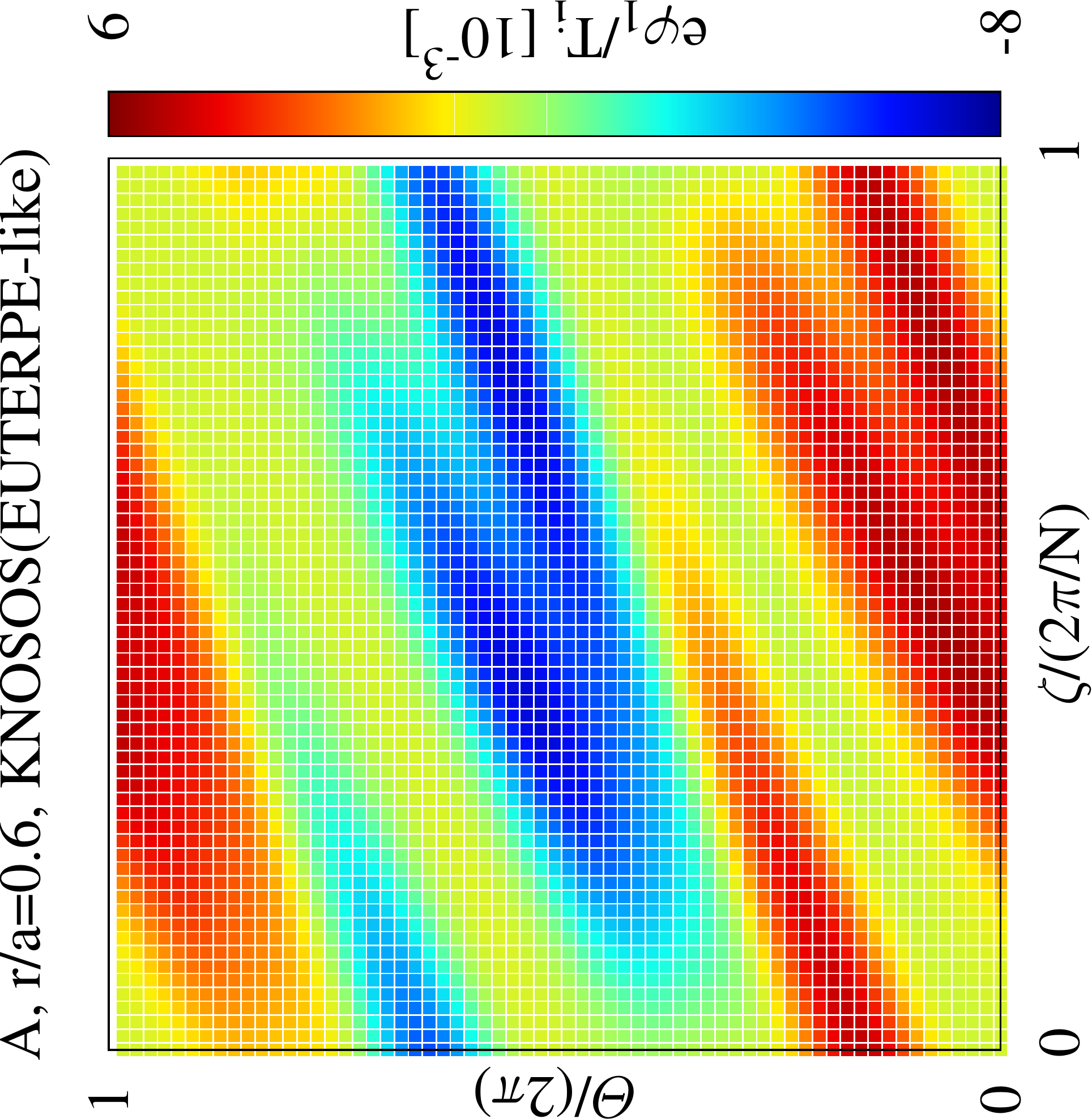}
\includegraphics[width=0.3\columnwidth,angle=270]{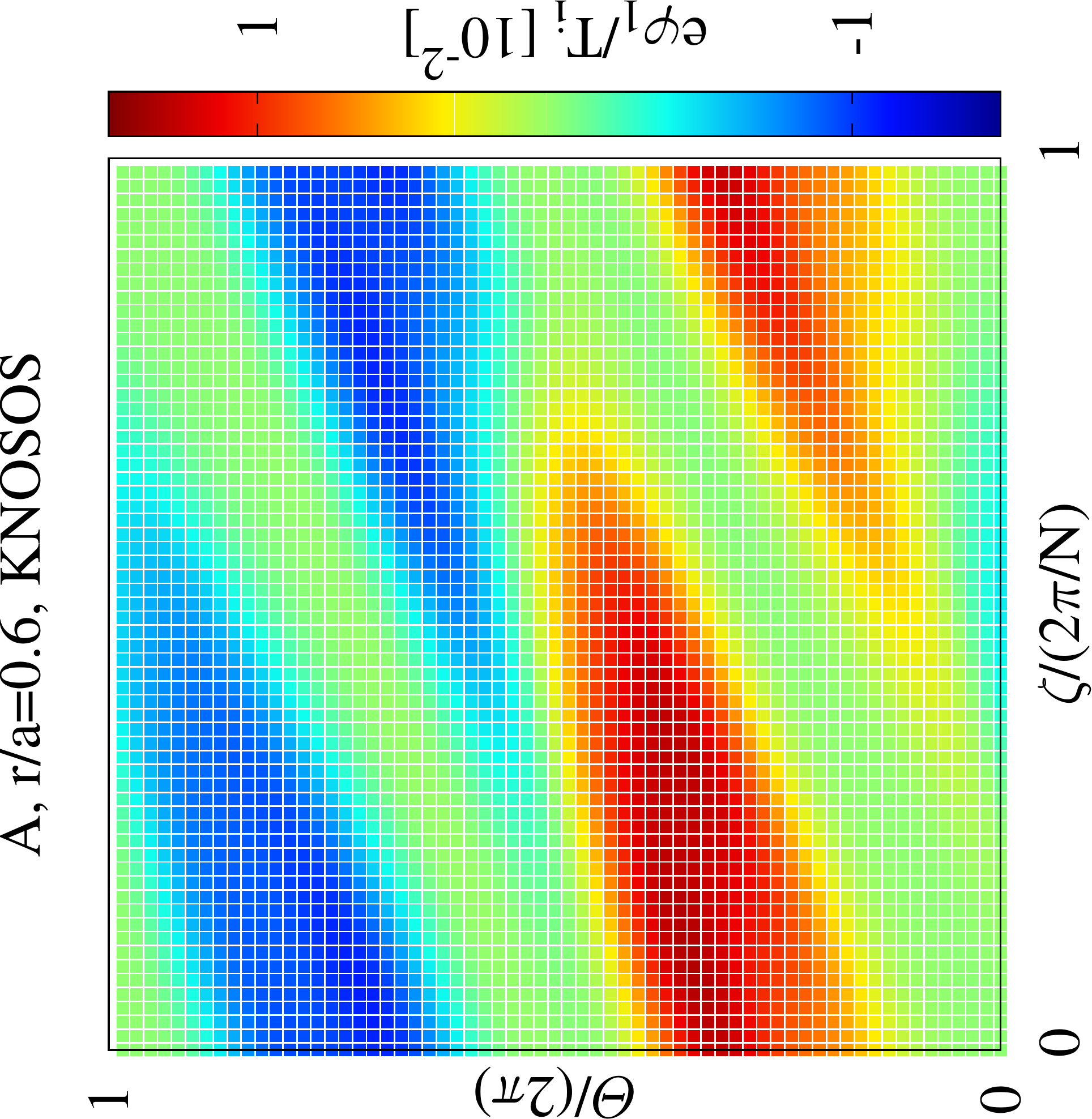}

\

\includegraphics[width=0.3\columnwidth,angle=270]{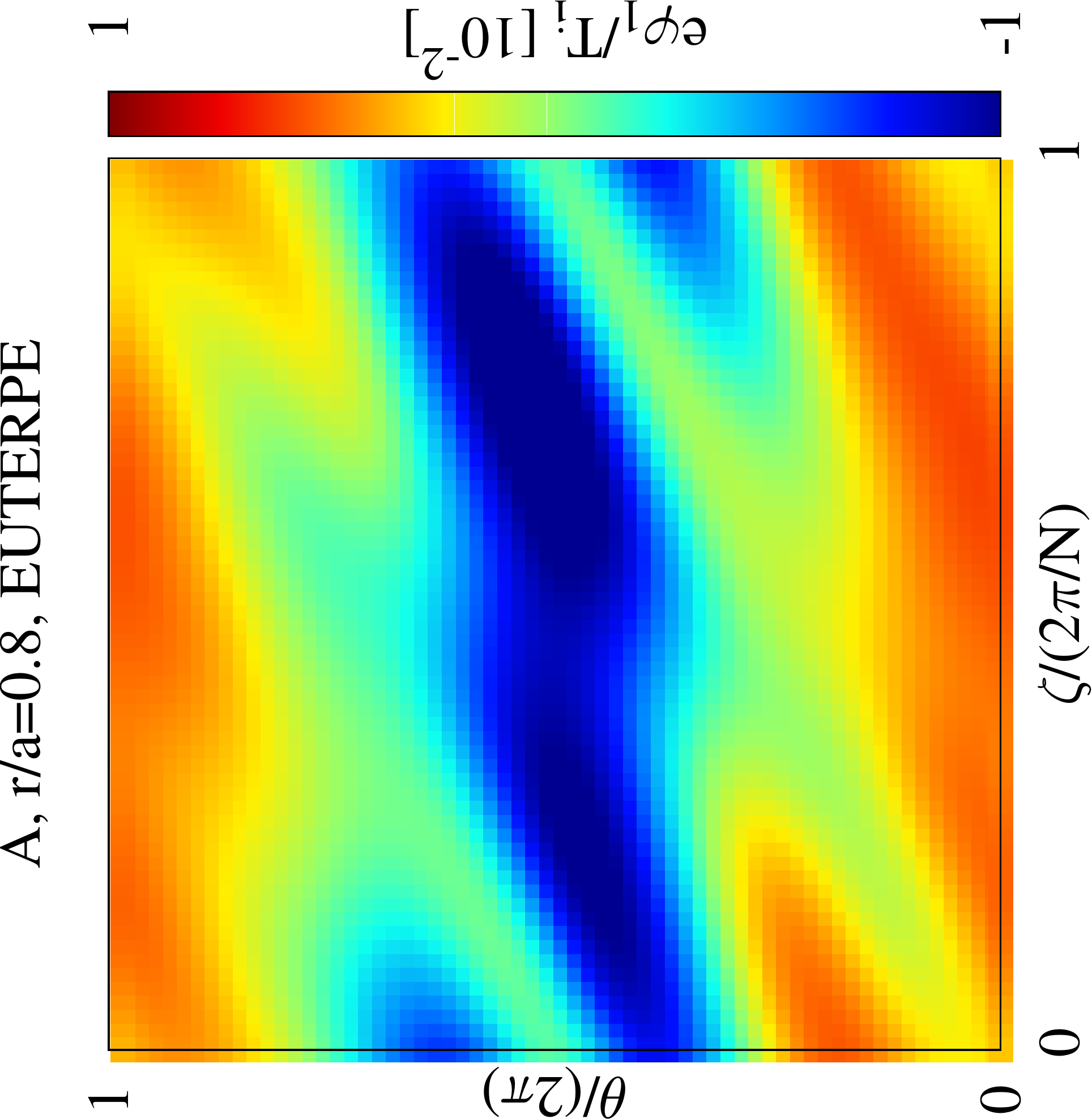}
\includegraphics[width=0.3\columnwidth,angle=270]{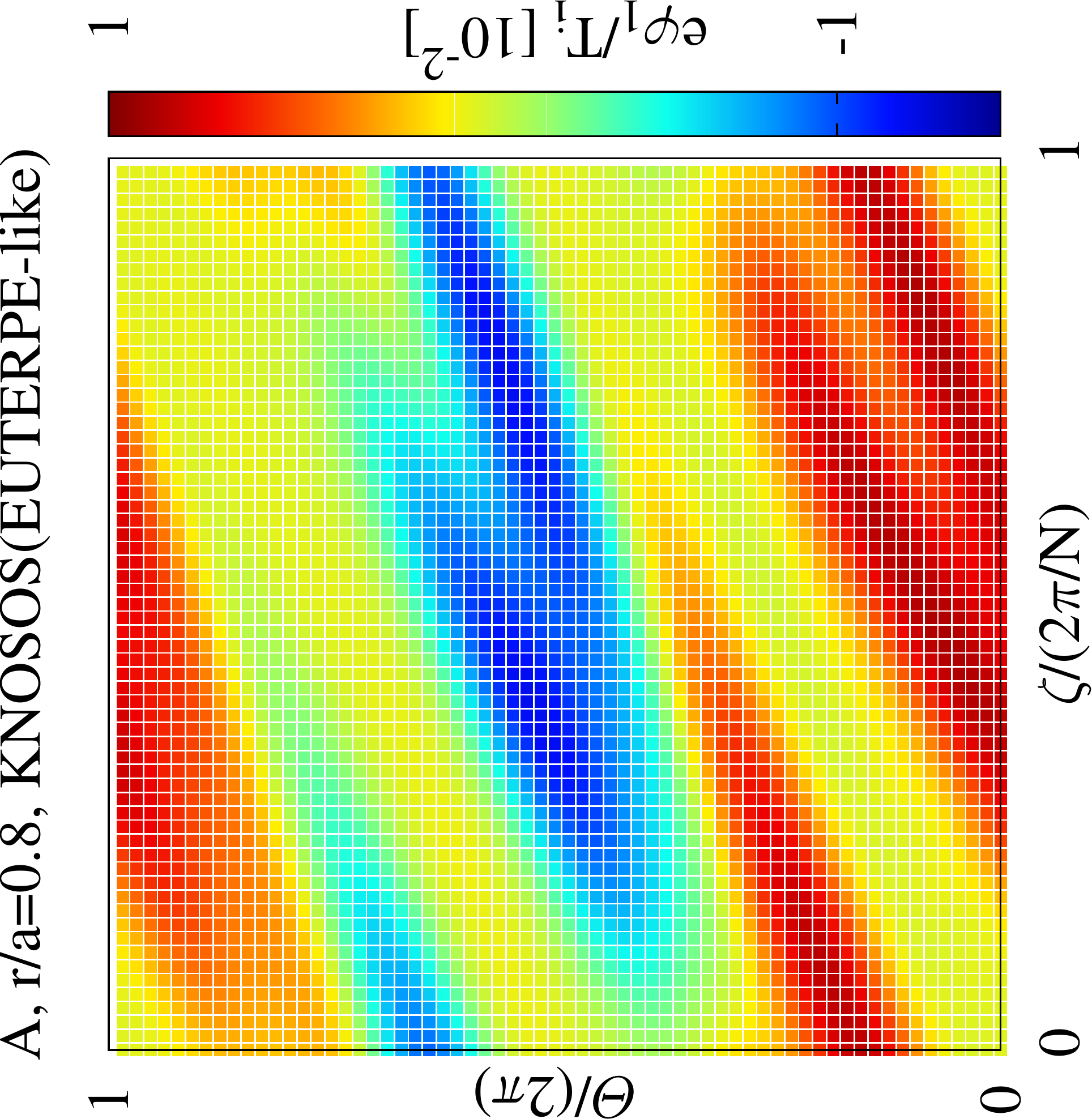}
\includegraphics[width=0.3\columnwidth,angle=270]{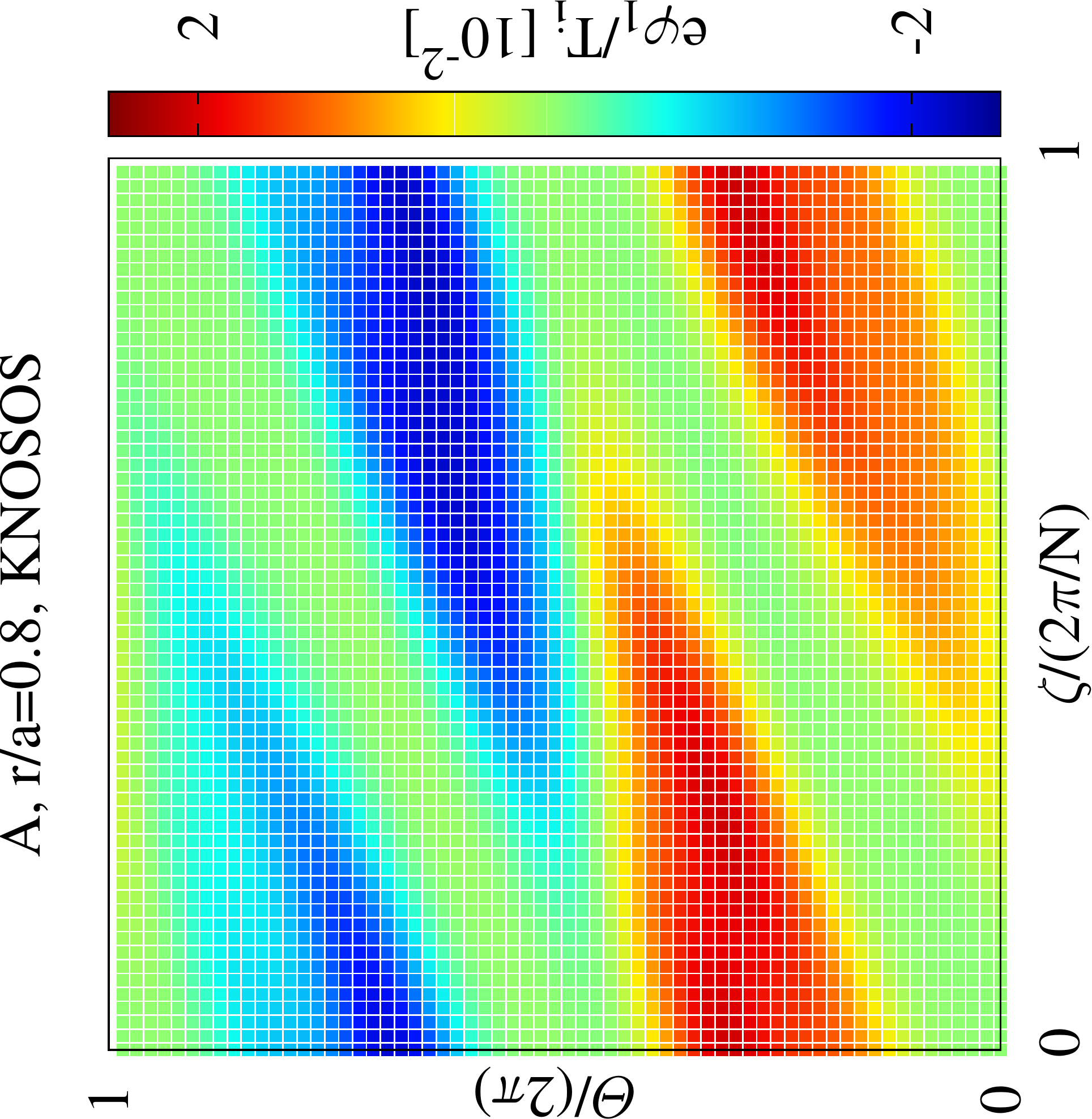}
\end{center}
\caption{Electrostatic potential variations on the flux surface calculated fort the plasma A with \texttt{EUTERPE} (left) and \texttt{KNOSOS} neglecting (center) and including (right) the tangential magnetic drift. The four rows correspond to radial positions $r/a\,=\,$0.2, 0.4, 0.6 and 0.8.}
\label{FIG_PHI1A3}
\end{figure}

\begin{figure}
\begin{center}
\includegraphics[width=\columnwidth,angle=0]{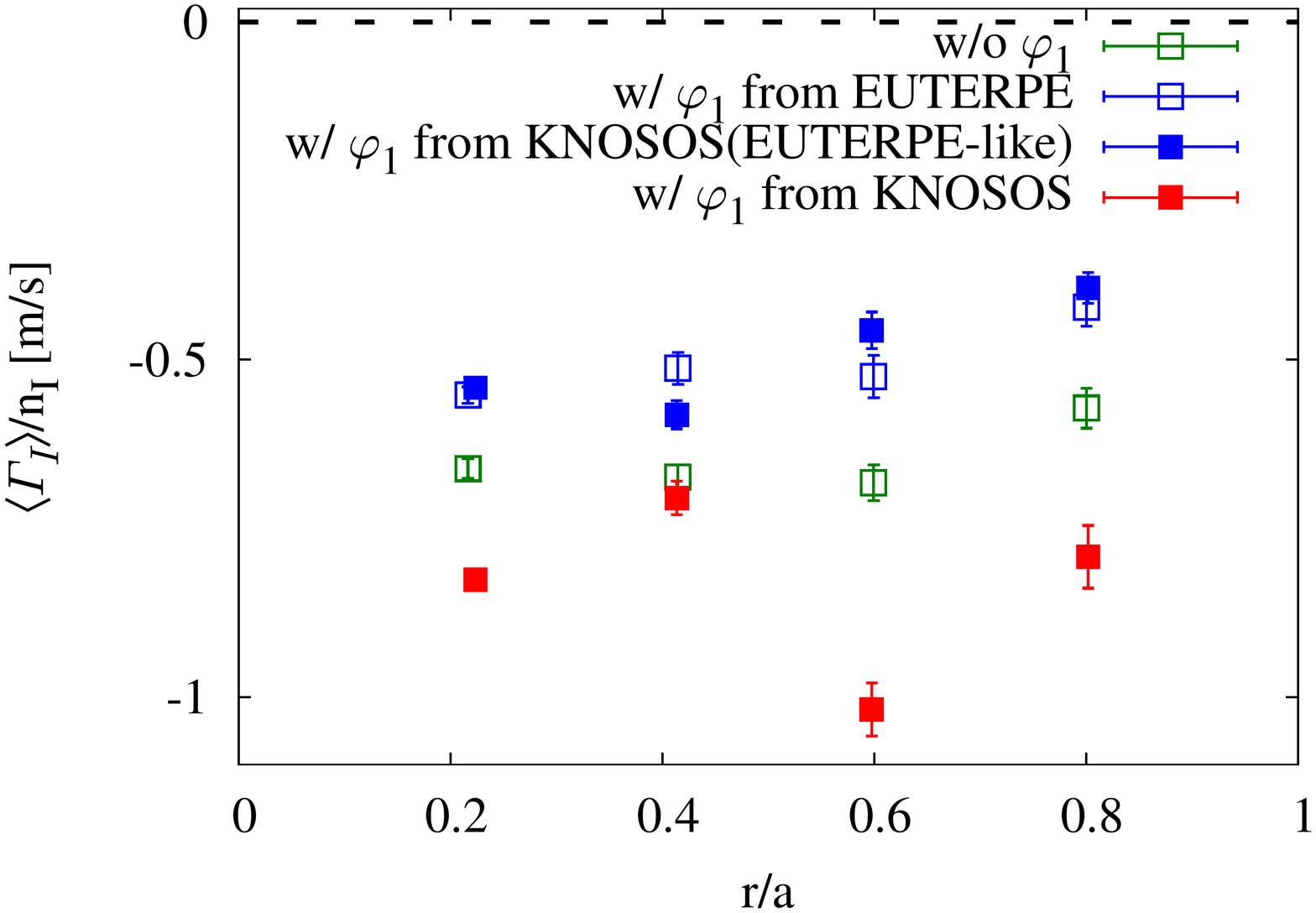}
\end{center}
\caption{Normalized radial flux of carbon for the plasma A.}
\label{FIG_FLUXA3}
\end{figure}

\begin{figure}
\begin{center}
\includegraphics[width=0.3\columnwidth,angle=270]{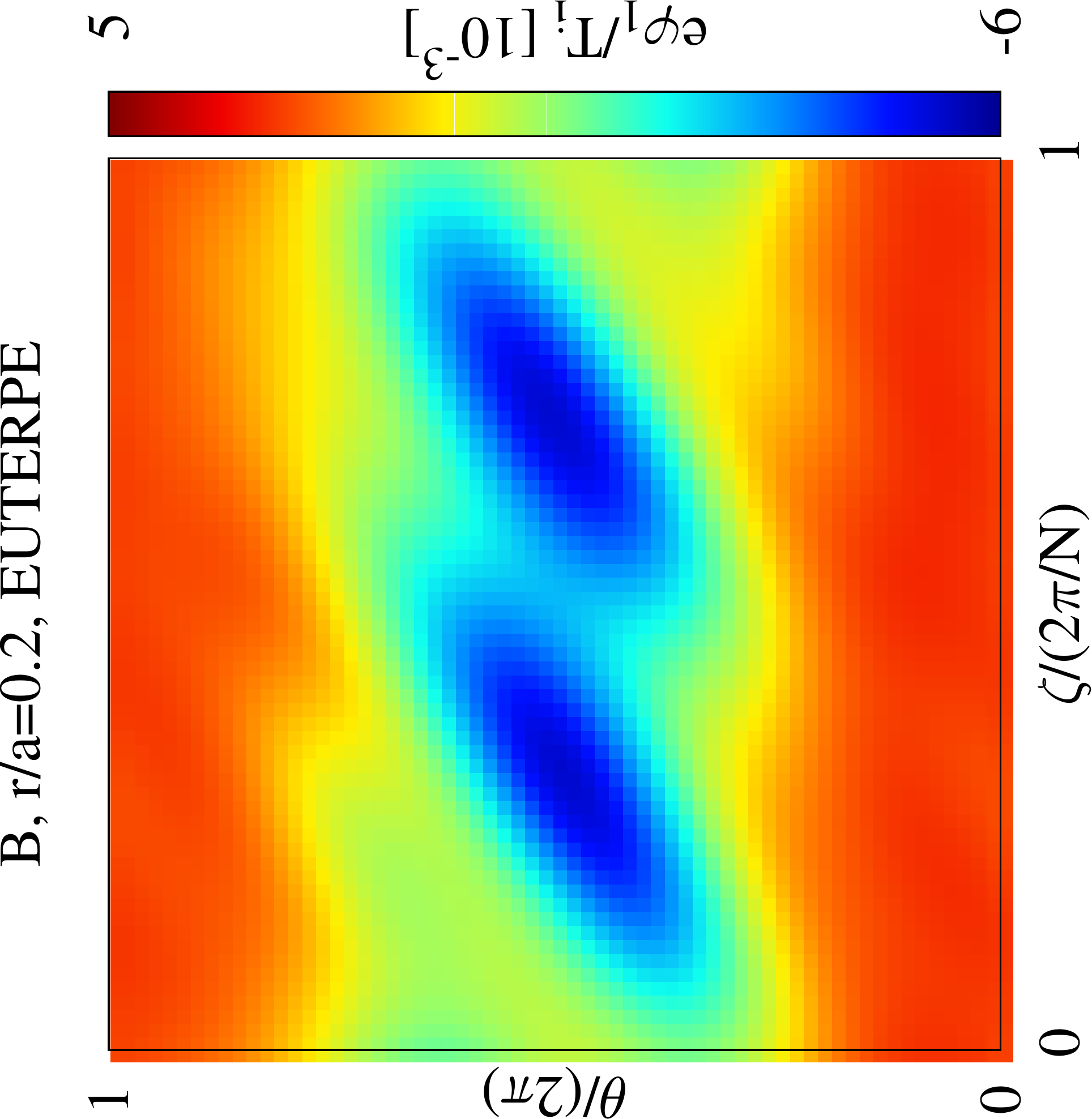}
\includegraphics[width=0.3\columnwidth,angle=270]{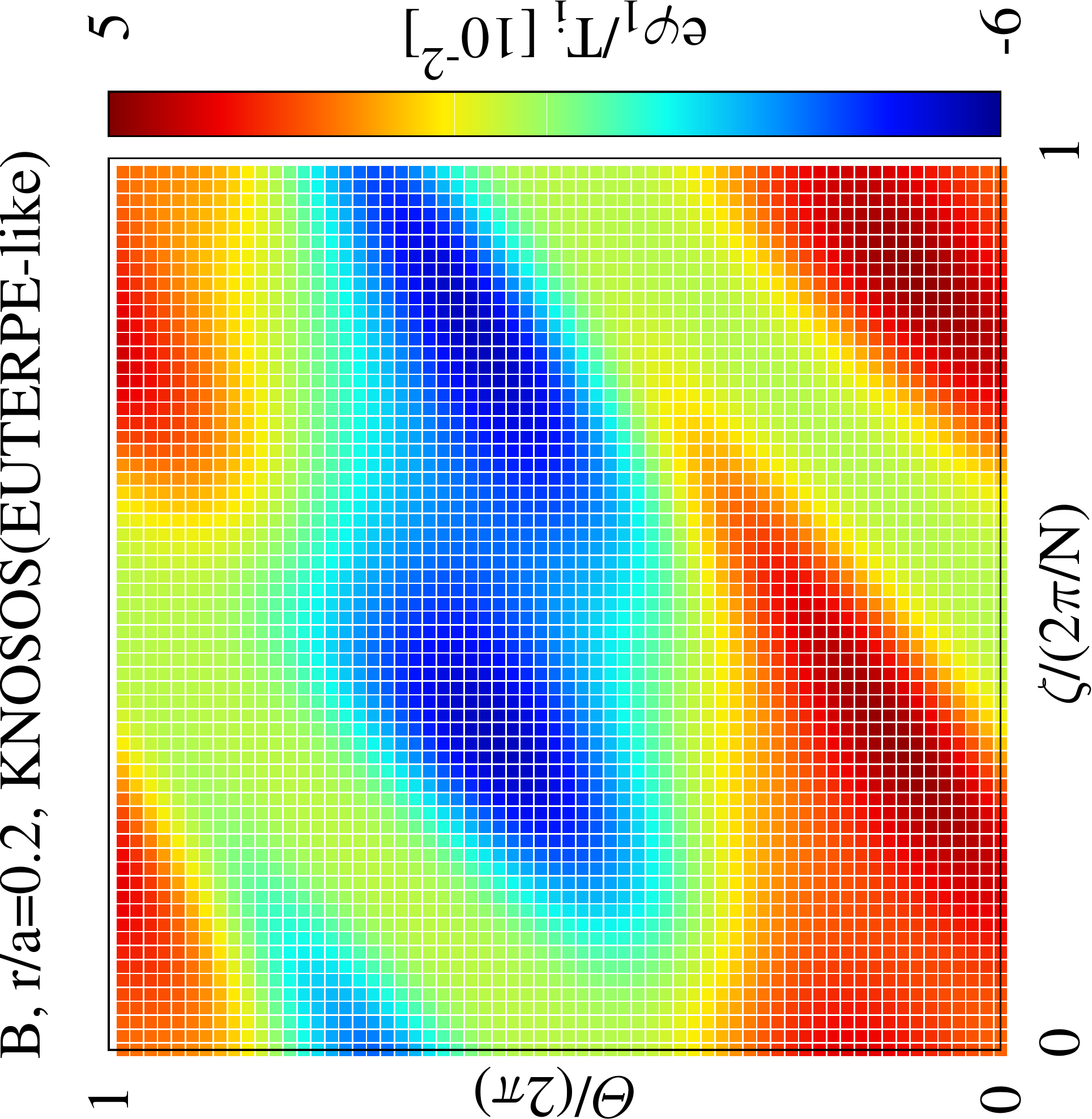}
\includegraphics[width=0.3\columnwidth,angle=270]{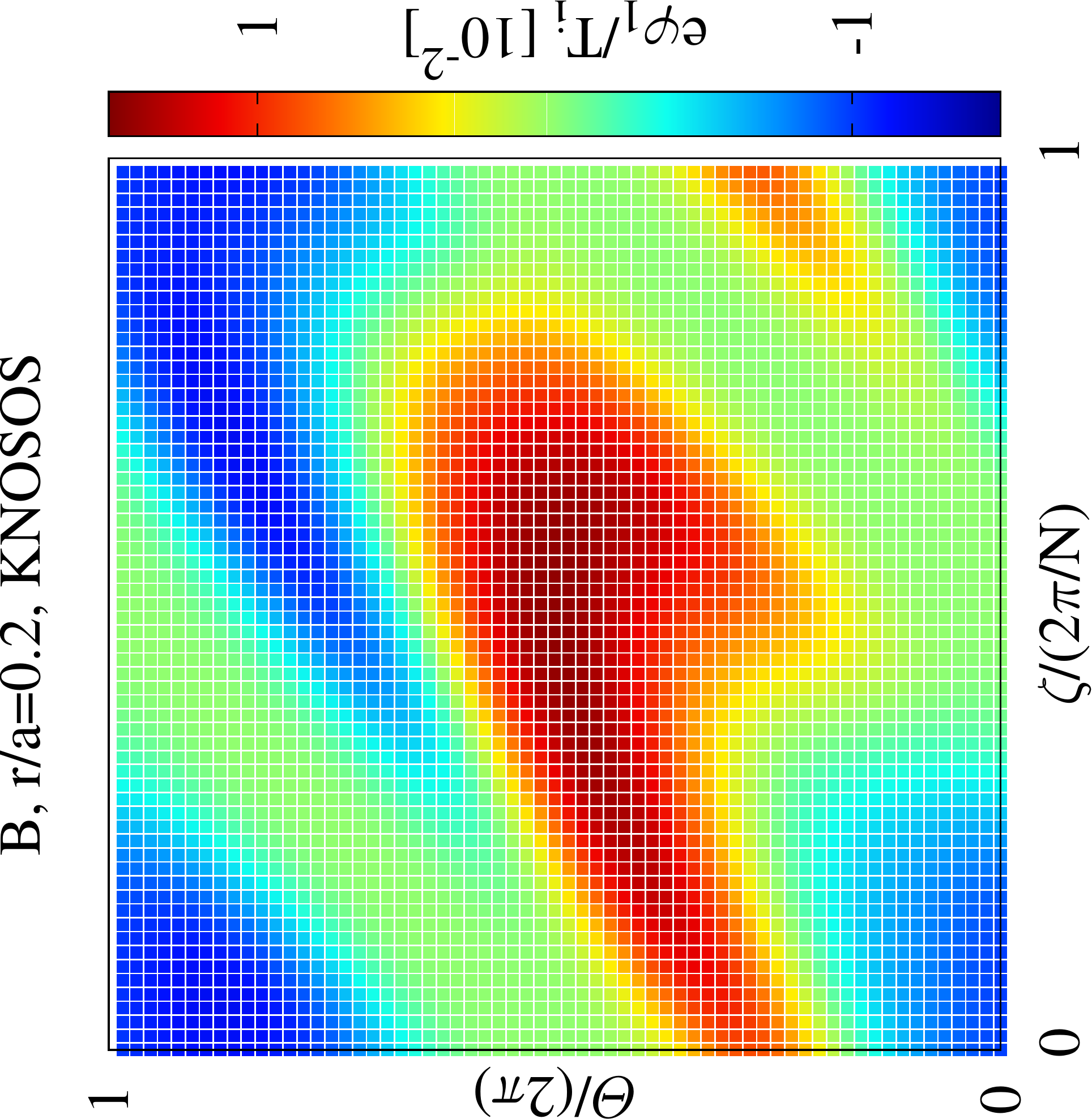}

\

\includegraphics[width=0.3\columnwidth,angle=270]{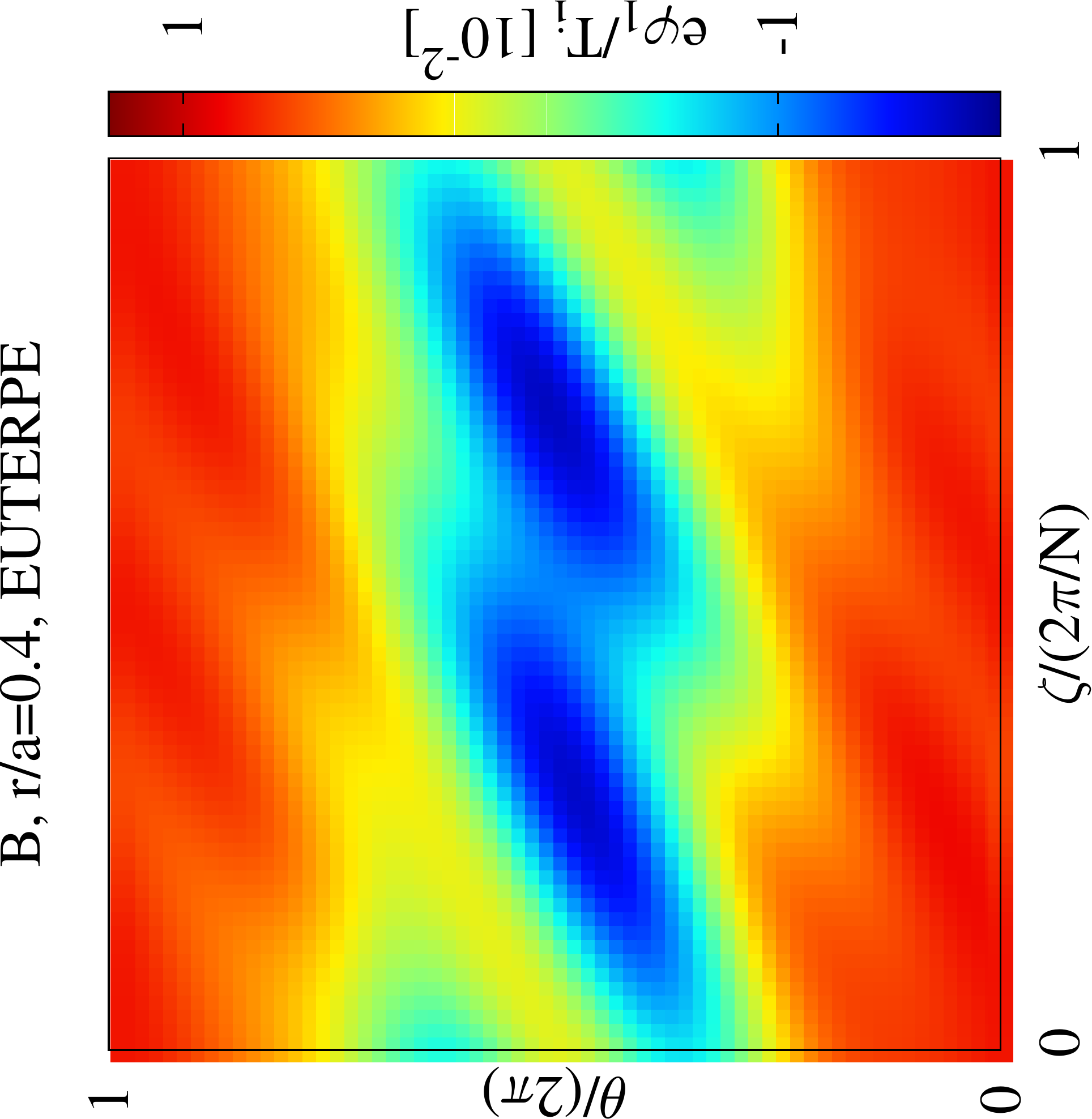}
\includegraphics[width=0.3\columnwidth,angle=270]{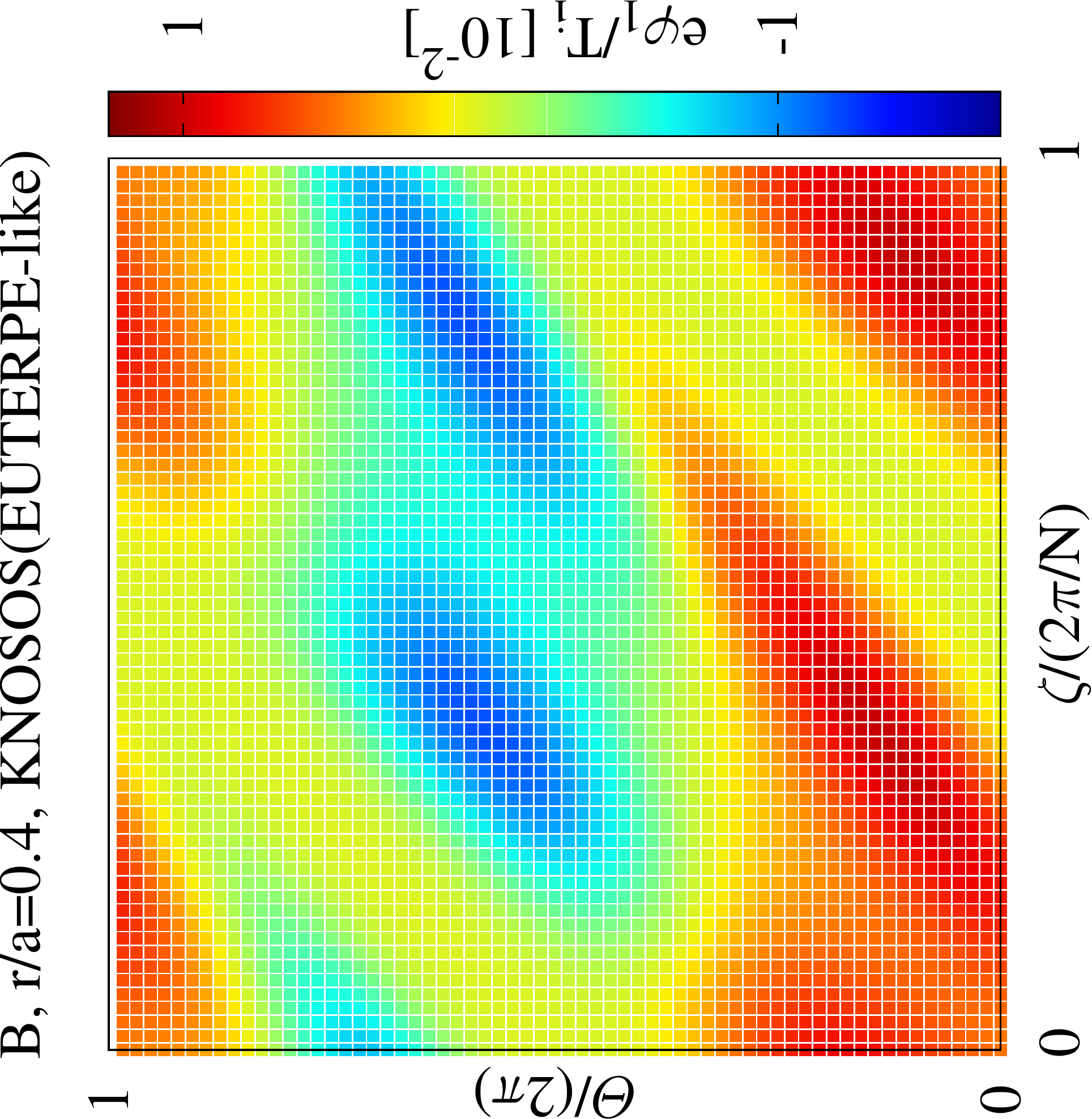}
\includegraphics[width=0.3\columnwidth,angle=270]{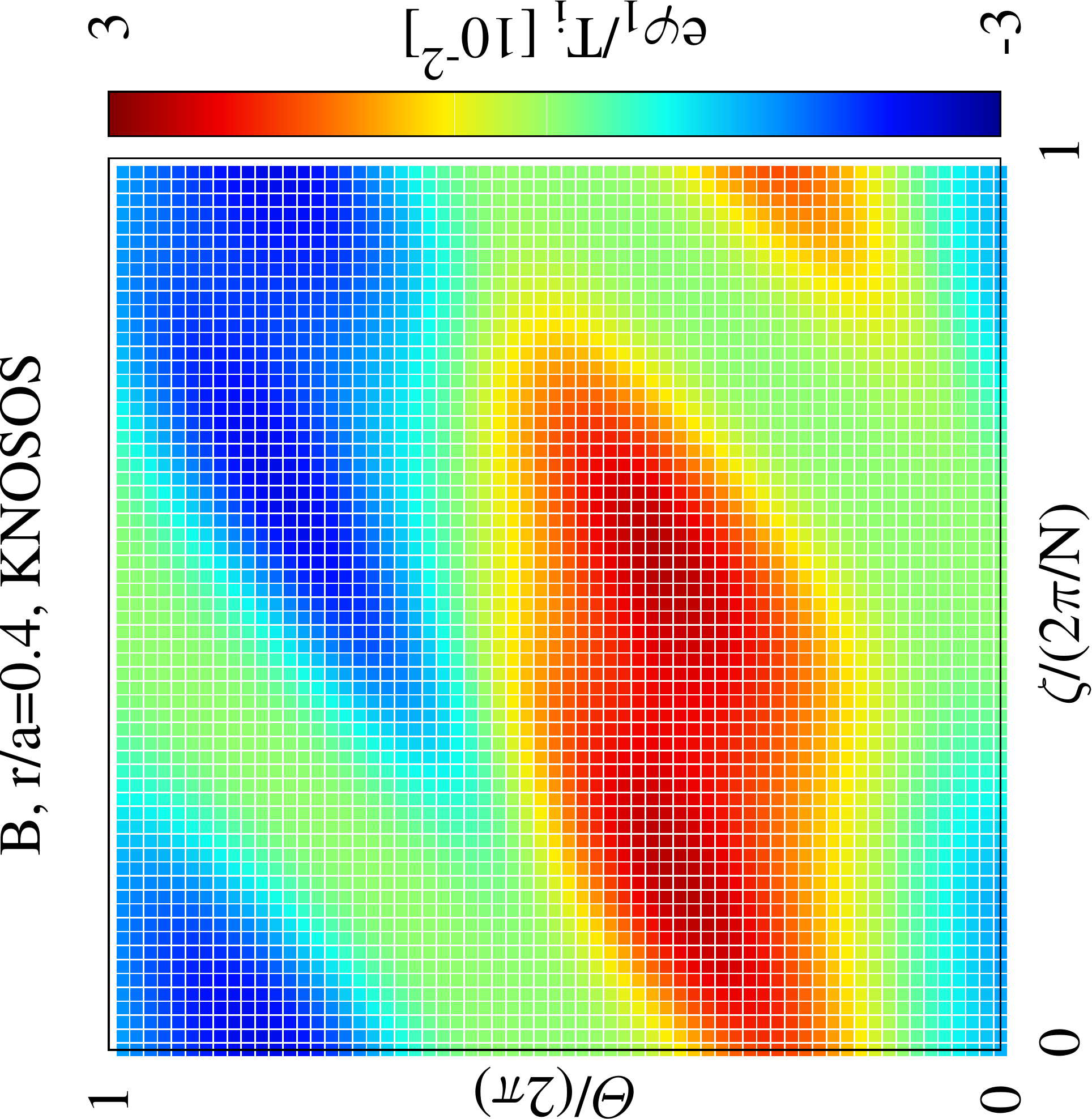}

\

\includegraphics[width=0.3\columnwidth,angle=270]{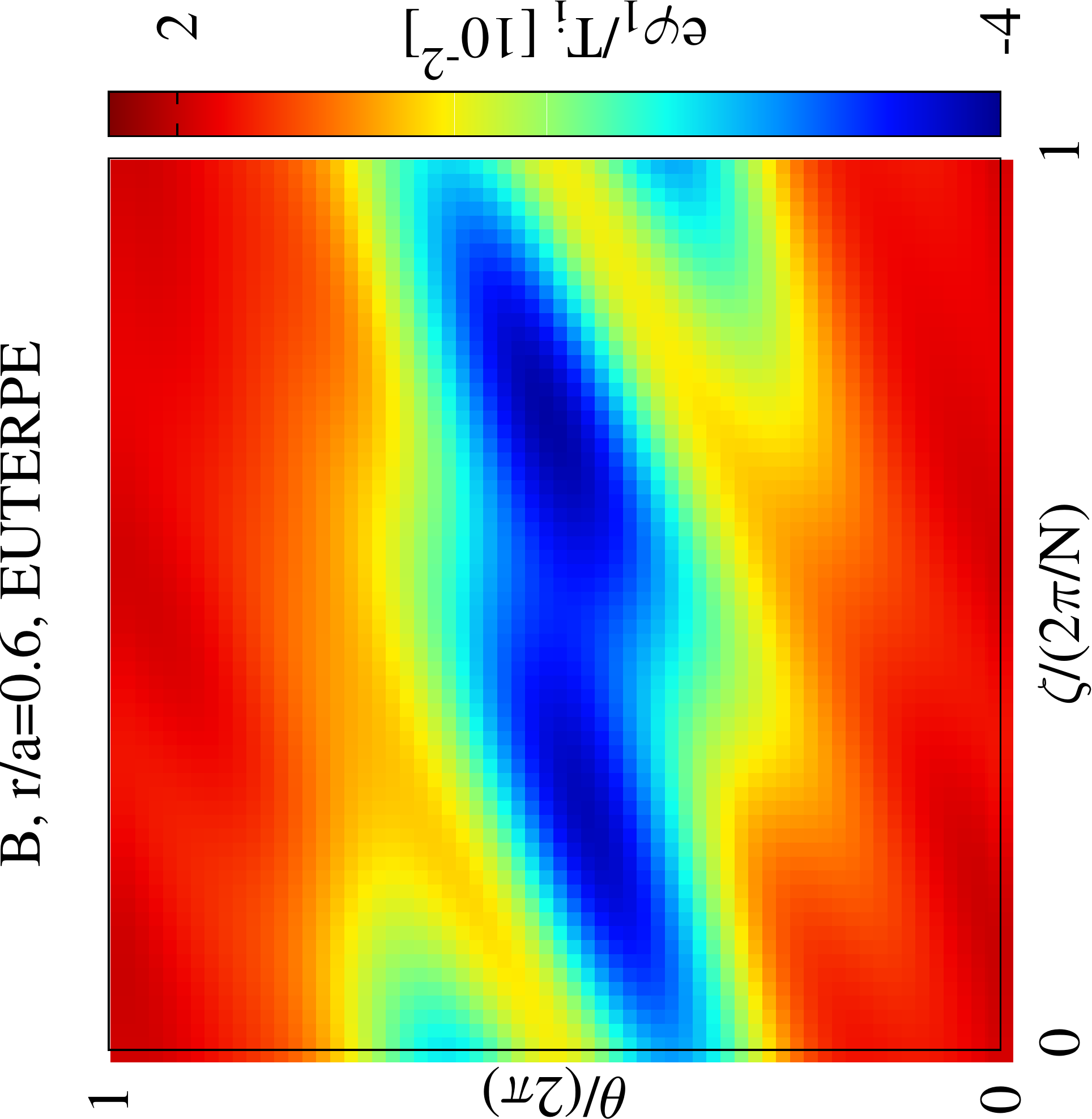}
\includegraphics[width=0.3\columnwidth,angle=270]{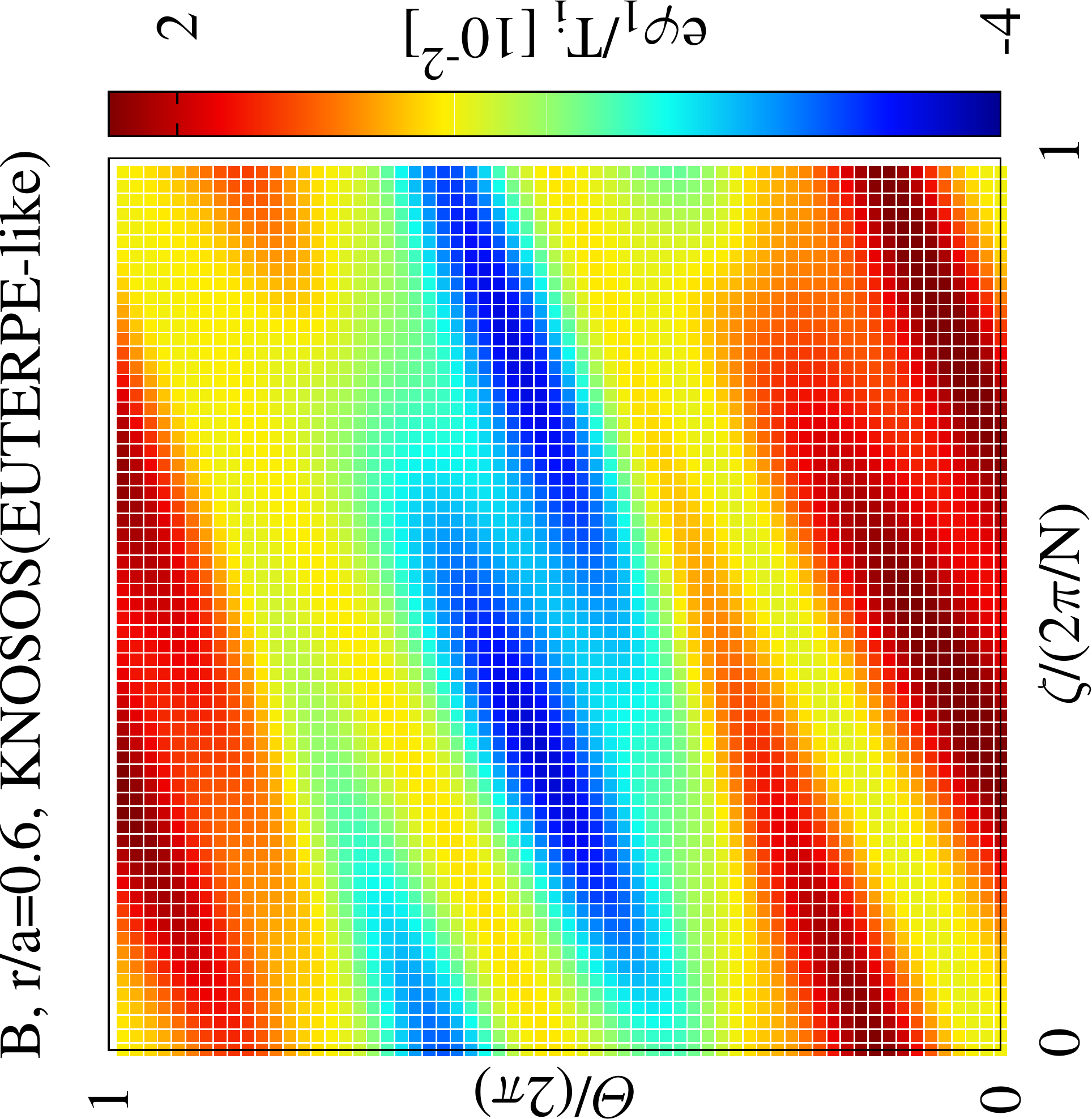}
\includegraphics[width=0.3\columnwidth,angle=270]{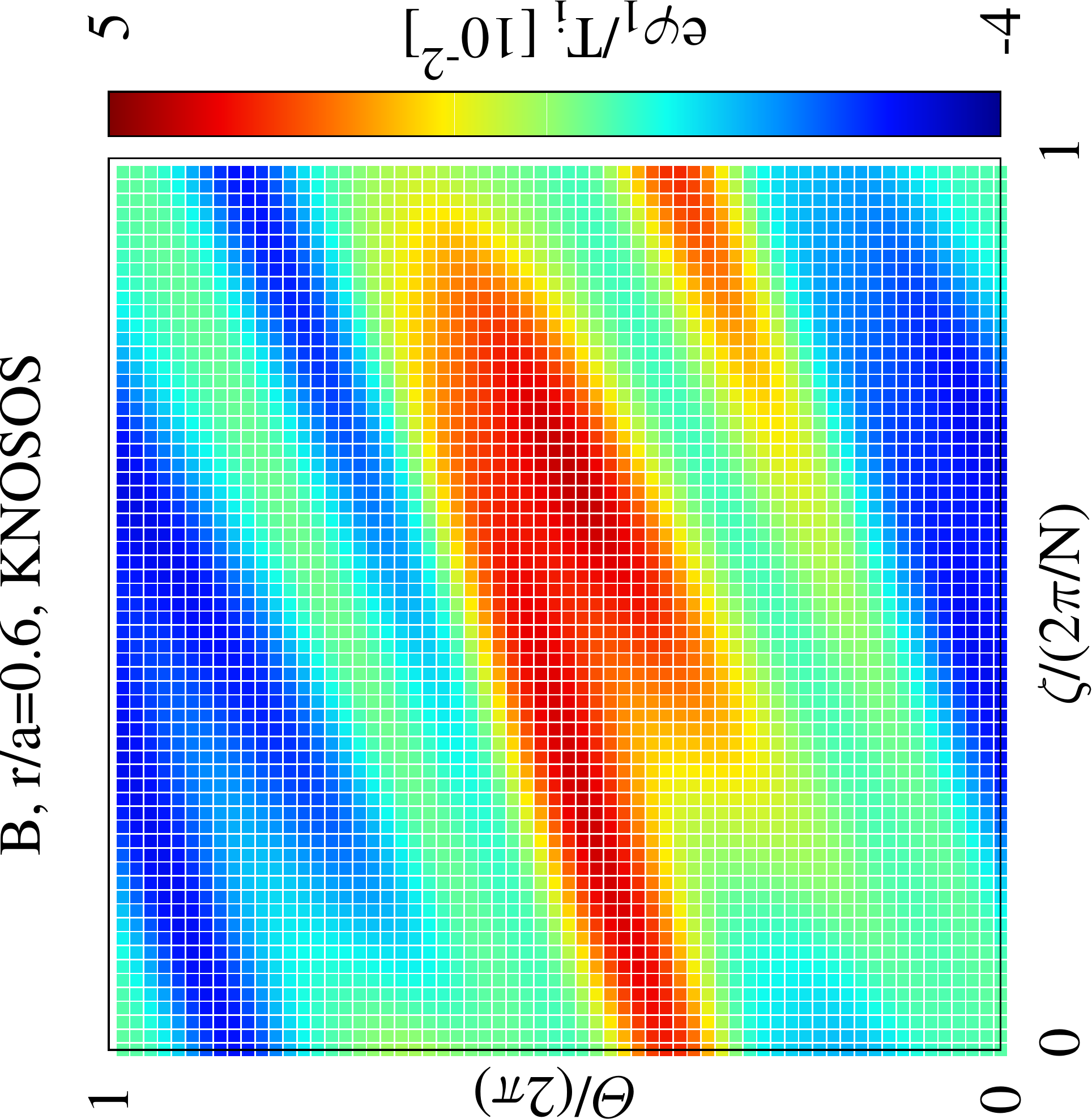}

\

\includegraphics[width=0.3\columnwidth,angle=270]{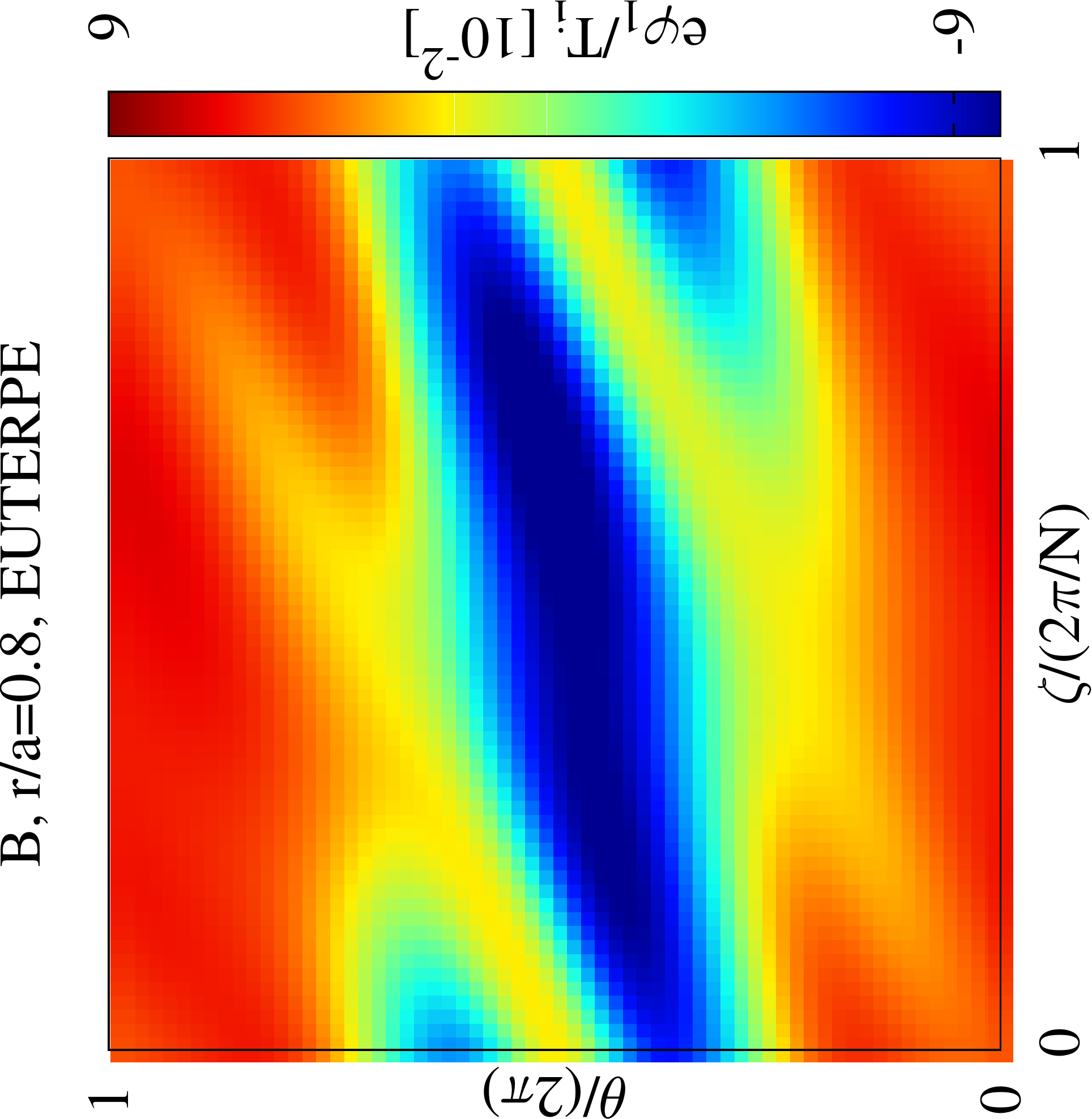}
\includegraphics[width=0.3\columnwidth,angle=270]{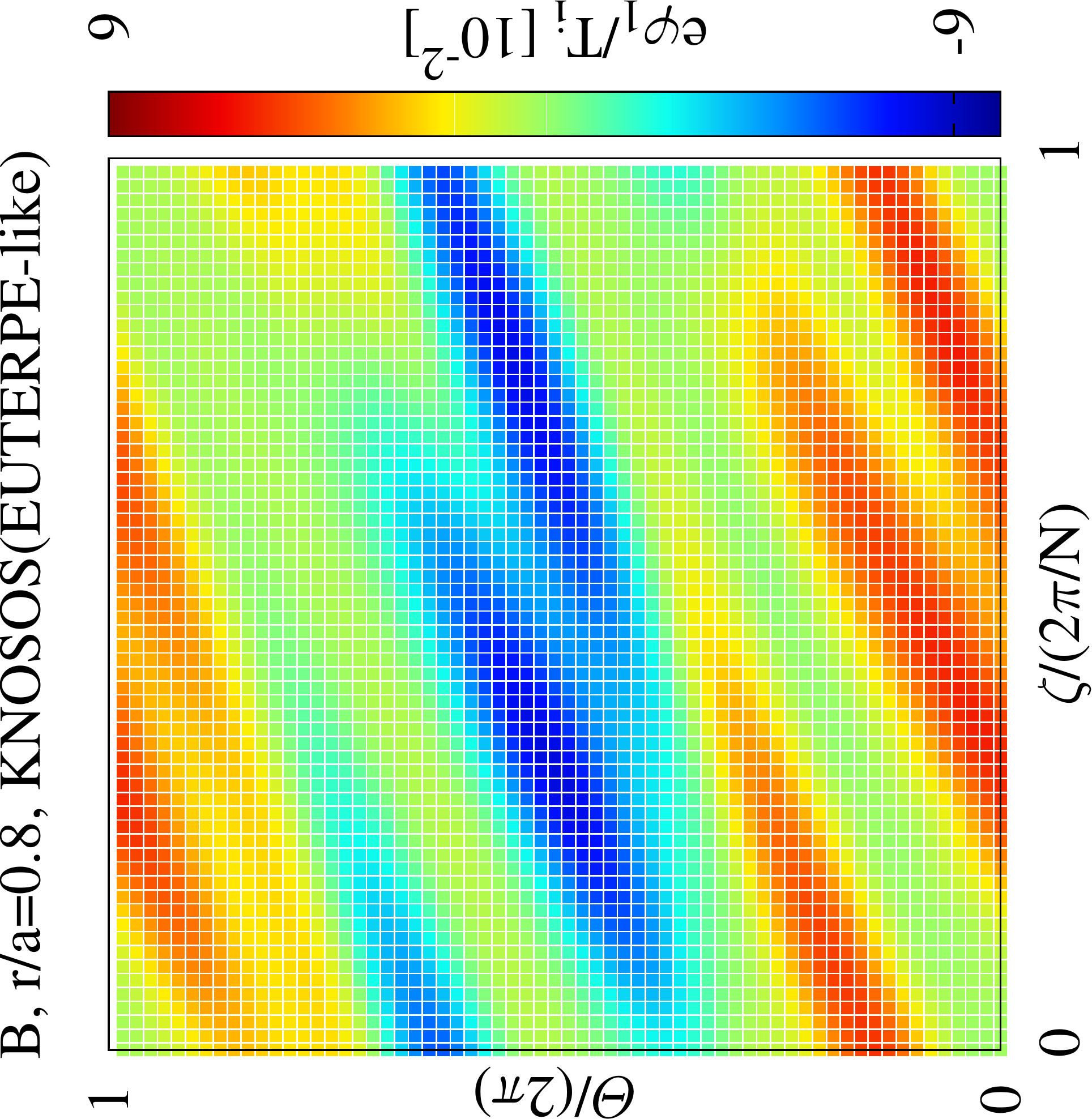}
\includegraphics[width=0.3\columnwidth,angle=270]{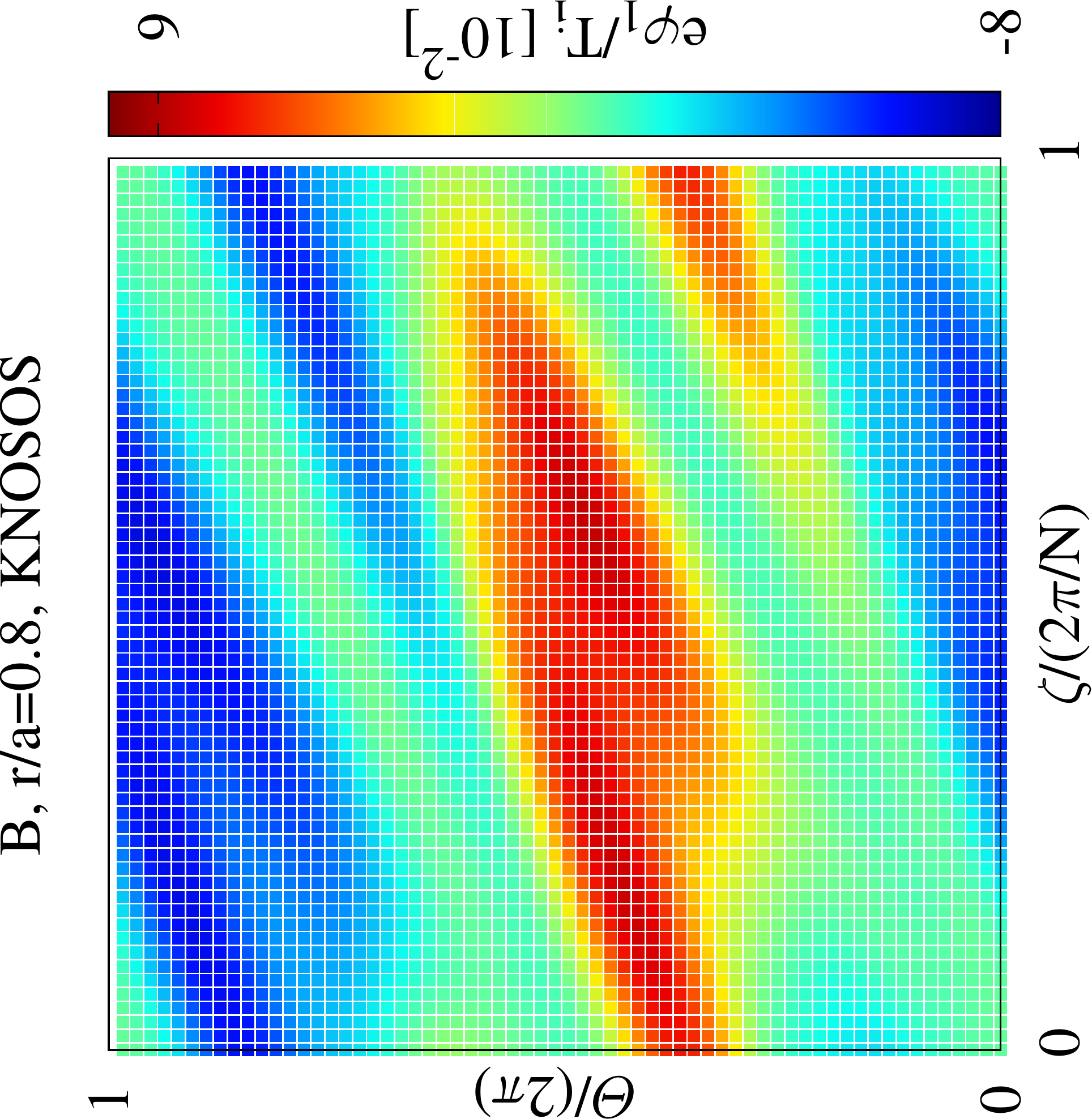}
\end{center}
\caption{Electrostatic potential variations on the flux surface calculated fort the plasma B with \texttt{EUTERPE} (left) and \texttt{KNOSOS} neglecting (center) and including (right) the tangential magnetic drift. The four rows correspond to radial positions $r/a\,=\,$0.2, 0.4, 0.6 and 0.8.}
\label{FIG_PHI1B3}
\end{figure}

\begin{figure}
\begin{center}
\includegraphics[width=\columnwidth,angle=0]{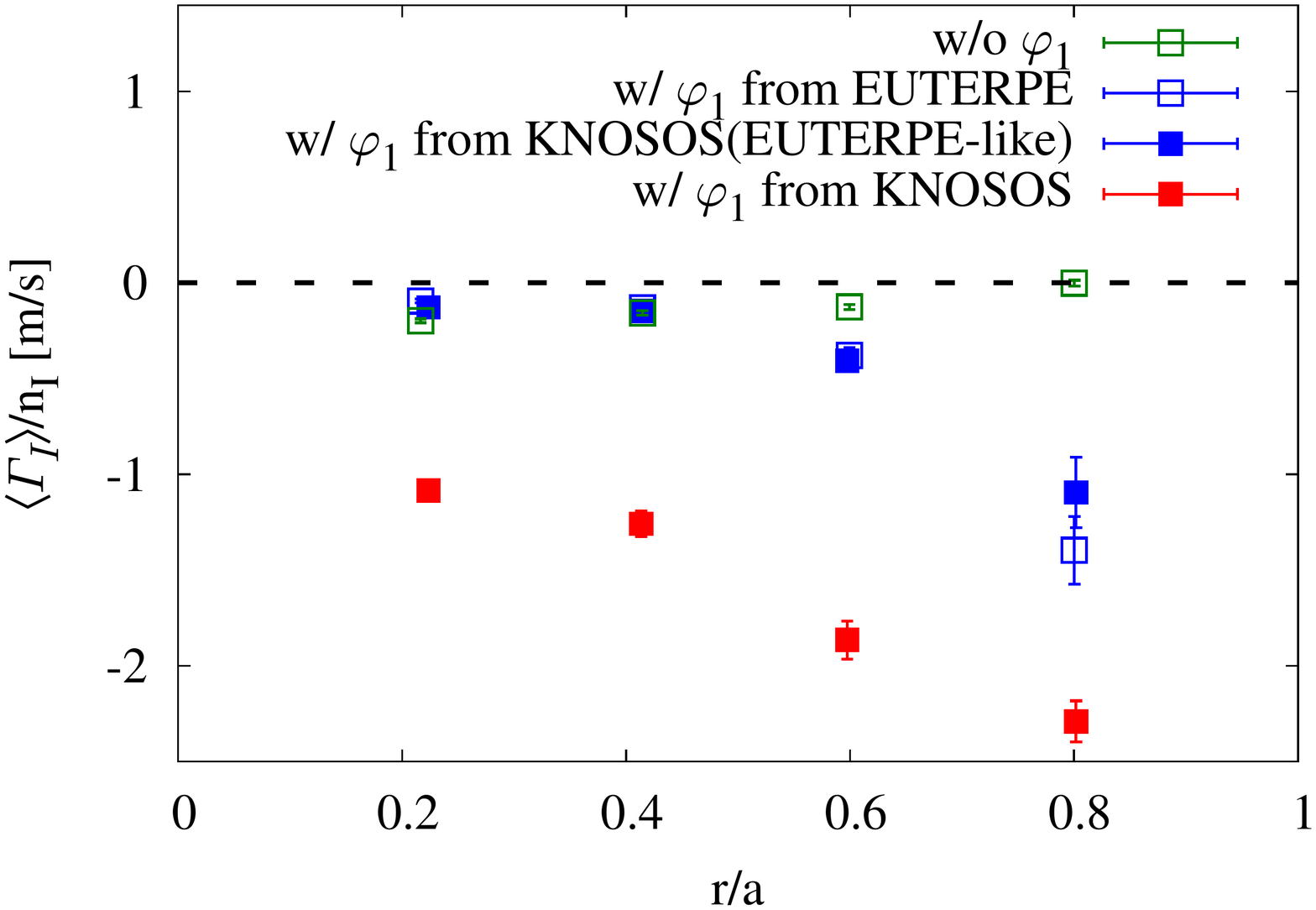}
\end{center}
\caption{Normalized radial flux of carbon for the plasma B.}
\label{FIG_FLUXB3}
\end{figure}

\subsection{Results}\label{SEC_RESULTS}

Figure~\ref{FIG_PHI1A3} shows the variation of the electrostatic potential on the flux surface for the plasma A of figure~\ref{FIG_PROFILES_PHI1}. Each row corresponds to a different flux surface, and each column to a different calculation method. Let us start by comparing the left column, calculated with \texttt{EUTERPE}, with the central column, calculated with \texttt{KNOSOS} without tangential magnetic drift (this is what we \ivan{call ``\texttt{KNOSOS} (\texttt{EUTERPE}-like)'')}. As long as the approximations in \texttt{KNOSOS} hold, the two methods should give the same results, and we observe that, although there are some differences, reasonable agreement between the two codes is obtained: the amplitude of $\varphi_1$ is the same, and so is the phase. As expected for ions mainly in the $\sqrt{\nu}$ regime, the electrostatic potential has a large stellarator-symmetric component~\cite{calvo2017viena}. In both cases cases the maxima lie in a broad strip around $\Theta =\pi$ and the minima around $\Theta=0$ and $\Theta =2\pi$. Only at the innermost positions, where the radial electric field is small, the stellerator-antisymmetric component in $\varphi_1$ becomes prominent. \texttt{KNOSOS} is able to capture the deviation from stellarator-symmetry (even in the clearly stellerator-symmetric cases, see e.g.~$r/a=0.6$, one can notice that the minimum of $\varphi_1$ actually lies at $\Theta\gtrsim\pi$), and also the structure of two separate minima around $\Theta =\pi$ and two separate maxima around $\Theta =0$. The main differences come from the angular region where $B$ and $B_0$ take their maximum values at each flux surface. \ivan{In that region} the contribution of the passing particles, which is typically \ivan{small,} may become non-negligible \ivan{because the contribution} of the trapped particles goes to zero, as the integration over $\lambda$ in equation~(\ref{EQ_QN}) is \ivan{taken over} a shrinking domain.

Before proceeding to the complete calculations, it is worth checking if these small differences between \texttt{EUTERPE} and \texttt{KNOSOS} (\texttt{EUTERPE}-like) are translated into inaccuracies in the computation of the radial impurity flux. In order to do so, we calculate \jlvg{with \texttt{EUTERPE} the flux of fully-stripped carbon ($Z_I=6$) in the trace-impurity limit using $\varphi_1$ from figure~\ref{FIG_PHI1A3}}. The results are shown in figure~\ref{FIG_FLUXA3}: where we see that \texttt{EUTERPE} and \texttt{KNOSOS} (\texttt{EUTERPE}-like) predict radial fluxes that are quite close, and clearly different to the calculation that does not consider the tangential electric field.

We are now in the position of discussing the complete calculation including the tangential magnetic drift in the solution of the drift-kinetic equation of the bulk ions. We see in figure~\ref{FIG_PHI1A3} (right) important differences with respect to figure~\ref{FIG_PHI1A3} (center): first of all the amplitude once the tangential magnetic drift is included is larger, more than a factor 2. Secondly, the phase changes: while the maps of figure~\ref{FIG_PHI1A3} (center) for $r/a>0.2$ are mainly stellarator-symmetric (as expected for ions in the $\sqrt{\nu}$ regime), once we include the tangential magnetic drift and the superbanana-plateau regime shows up, the map has no definite symmetry. \jlvg{Once we have outlined these clear differences, we do not describe the results in more detail since, as we have discussed, we do not expect these simulations to be quantitatively accurate, and we thus turn our attention to impurity transport.} 

The effect of the tangential electric field on the impurity flux is given by its amplitude and by the relative phase of the electrostatic potential variations with respect to the variation of the impurity density on the flux surface. Therefore, the effect discussed in the previous paragraph should have an impact on the impurity flux. We indeed see in figure~\ref{FIG_FLUXA3} that including the tangential magnetic drift changes the sign of the contribution of $\varphi_1$ for the plasmas of A: the radial flux becomes more negative, being specially large at $r/a=0.6$.

Finally, we \jlvg{briefly repeat the discussion of the two previous paragraphs for} plasma B of figure~\ref{FIG_PROFILES_PHI1}\jlvg{, keeping the same caveats in mind}. This is a plasma of lower collisionality and smaller absolute value of the radial electric field, and we thus expect $\varphi_1$ to be larger ( let us remember that our linearization relies on $\varphi_1$ not becoming too large). In figure~\ref{FIG_PHI1B3} we show the variation of the electrostatic potential calculated, as in figure~\ref{FIG_PHI1A3}, for four radial positions using three different methods. The agreement between \texttt{EUTERPE} and \texttt{KNOSOS} (\texttt{EUTERPE}-like) remains reasonably good, and the same conclusion can be drawn for the corresponding calculations of radial impurity flux shown in figure~\ref{FIG_FLUXB3}. We see that including the tangential magnetic drift in the calculation of $\varphi_1$ modifies the result completely. Apart from the change in amplitude and phase, the more salient result is that a thin strip of extreme values of $\varphi_1$ starts to develop, with the same helicity of the underlying $B_0$, i.e. $\varphi_1=\varphi_1(\Theta-N\zeta)$, at $r/a<0.5$ and $\varphi_1=\varphi_1(2\Theta-N\zeta)$, at $r/a>0.5$, being most visible at $r/a=0.6$ and 0.8. This is expected to come from the contribution of ions in the superbanana-plateau for small enough radial electric field, as discussed in~\cite{calvo2017sqrtnu} (detailed simulations for this particular regime were presented in~\cite{calvo2017viena}). The effect on the impurity flux of the superbanana-plateau contribution to $\varphi_1$ is expected to be even larger than in the plasma A. In figure~\ref{FIG_FLUXB3} we show indeed than the carbon flux becomes very negative.

\section{Conclusions}\label{SEC_CONCLUSIONS}

In this paper it has been shown that plasmas of low collisionality are closer to impurity screening than previously expected, as the radial electric field becomes smaller than the ion temperature gradient. In these situation, large tangential electric drifts had been predicted coming from the superbanana-plateau regime of the bulk ions. Therefore, we have solved the quasineutrality equation and the drift-kinetic equation of the bulk ions including for the first time the effect of the tangential magnetic drift, using the newly developed code \texttt{KNOSOS}. We have shown that this contribution may be very relevant for the impurity transport of realistic plasmas such as those displaying impurity hole, and we have argued that that radially-global simulations are needed to compute it accurately for experimentally-relevant plasmas.

\section{Acknowledgments}

This work has been carried out within the framework of the EUROfusion Consortium and has received funding from the Euratom research and training programme 2014-2018 under grant agreement No 633053. The views and opinions expressed herein do not necessarily reflect those of the European Commission. The International Stellarator-Heliotron Database is pursued under the auspices of IEA Implementing Agreement for Cooperation in Development of the Stellarator/­Heliotron Concept (2.10.1992) and the authors are indebted to all its contributors and previous responsible officers. This research was supported in part by grant ENE2015-70142-P, Ministerio de Econom\'ia y Competitividad, Spain.

\section{Bibliography}


\begin{thebibliography}{10}

\bibitem{ida2009observation}
K~Ida, M~Yoshinuma, M~Osakabe, K~Nagaoka, M~Yokoyama, H~Funaba, C~Suzuki,
  T~Ido, A~Shimizu, I~Murakami, N~Tamura, H~Kasahara, Y~Takeiri, K~Ikeda,
  K~Tsumori, O~Kaneko, S~Morita, M~Goto, K~Tanaka, K~Narihara, T~Minami,
  I~Yamada, and {the LHD Experimental Group}.
\newblock {Observation of an impurity hole in a plasma with an ion internal
  transport barrier in the Large Helical Device}.
\newblock {\em Physics of Plasmas}, 16(5):056111, 2009.

\bibitem{yoshinuma2009observation}
M~Yoshinuma, K~Ida, M~Yokoyama, M~Osakabe, K~Nagaoka, S~Morita, M~Goto,
  N~Tamura, C~Suzuki, S~Yoshimura, H~Funaba, Y~Takeiri, K~Ikeda, K~Tsumori,
  O~Kaneko, and {the LHD Experimental Group}.
\newblock {Observation of an impurity hole in the Large Helical Device}.
\newblock {\em Nuclear Fusion}, 49(6):062002, 2009.

\bibitem{mccormick2002hdh}
K~McCormick, P~Grigull, R~Burhenn, R~Brakel, H~Ehmler, Y~Feng, F~Gadelmeier,
  L~Giannone, D~Hildebrandt, M~Hirsch, R~Jaenicke, J~Kisslinger, T~Klinger,
  S~Klose, J~Knauer Pand~R K\"onig, G~K\"uhner, H~Laqua P, D~Naujoks,
  H~Niedermeyer, E~Pasch, N~Ramasubramanian, N~Rust, F~Sardei, F~Wagner,
  A~Weller, U~Wenzel, and A~Werner.
\newblock New advanced operational regime on the w7-as stellarator.
\newblock {\em Physical Review Letters}, 89:015001, 2002.

\bibitem{igitkhanov2006impurity}
Y~Igitkhanov, E~Polunovsky, and C~D Beidler.
\newblock {Impurity dynamics in nonaxisymmetric plasmas}.
\newblock {\em Fusion science and technology}, 50(2):268, 2006.

\bibitem{regana2013euterpe}
J~M Garc\'ia-Rega{\~n}a, R~Kleiber, C~D Beidler, Y~Turkin, H~Maassberg, and
  P~Helander.
\newblock {On neoclassical impurity transport in stellarator geometry}.
\newblock {\em Plasma Physics and Controlled Fusion}, 55(7):074008, 2013.

\bibitem{regana2017phi1}
J.M. Garc{\'\i}a-Rega{\~n}a, C.D. Beidler, R.~Kleiber, P.~Helander,
  A.~Moll{\'e}n, J.A. Alonso, M.~Landreman, H.~Maa{\ss}berg, H.M. Smith,
  Y.~Turkin, and J.L. Velasco.
\newblock Electrostatic potential variation on the flux surface and its impact
  on impurity transport.
\newblock {\em Nuclear Fusion}, 57(5):056004, 2017.

\bibitem{mikkelsen2014gk}
D~R Mikkelsen, K~Tanaka, M~Nunami, T~H Watanabe, H~Sugama, M~Yoshinuma, K~Ida,
  Y~Suzuki, M~Goto, S~Morita, B~Wieland, I~Yamada, R~Yasuhara, T~Tokuzawa,
  T~Akiyama, and N~A Pablant.
\newblock {Quasilinear carbon transport in an impurity hole plasma in LHD}.
\newblock {\em Physics of Plasmas}, 21(8):082302, 2014.

\bibitem{nunami2016iaea}
M~Nunami, M~Nakata, H~Sugama, M~Sato, J~L Velasco, S~Satake, and M~Yokoyama.
\newblock {Anomalous and neoclassical transport of hydrogen isotope and
  impurity ions in LHD plasmas}.
\newblock In {\em 26th IAEA Fusion Energy Conference, Kyoto, Japan}, 2016.

\bibitem{alonso2016inertia}
J~A Alonso, I~Calvo, T~Estrada, J~M Fontdecaba, J~Garc\'ia-Rega{\~n}a,
  J~Geiger, M~Landreman, K~McCarthy, F~Medina, B~Ph~Van Milligen, M~A Ochando,
  F~I Parra, and J~L Velasco.
\newblock {}.
\newblock {\em Plasma Physics and Controlled Fusion}, submitted, 2016.

\bibitem{helander2017prl}
P~Helander, S~L Newton, A~Moll\'en, and H~M Smith.
\newblock Impurity transport in a mixed-collisionality stellarator plasma.
\newblock {\em Phyics Review Letters}, 118:155002, Apr 2017.

\bibitem{dinklage2013ncval}
A~Dinklage, M~Yokoyama, K~Tanaka, J~L Velasco, D~L\'opez-Bruna, C~D Beidler,
  S~Satake, E~Ascas\'ibar, J~Ar\'evalo, J~Baldzuhn, Y~Feng, D~Gates, J~Geiger,
  K~Ida, M~Jakubowski, A~L\'opez-Fraguas, H~Maassberg, J~Miyazawa, T~Morisaki,
  S~Murakami, N~Pablant, S~Kobayashi, R~Seki, C~Suzuki, Y~Suzuki, Yu~Turkin,
  A~Wakasa, R~Wolf, H~Yamada, M~Yoshinuma, {LHD Exp. Group}, {TJ-II Team}, and
  {W7-AS Team}.
\newblock {Inter-machine validation study of neoclassical transport modelling
  in medium- to high-density stellarator-heliotron plasmas}.
\newblock {\em Nuclear Fusion}, 53(6):063022, 2013.

\bibitem{calvo2017sqrtnu}
I~Calvo, F~I Parra, J~L Velasco, and A~Alonso.
\newblock The effect of tangential drifts on neoclassical transport in
  stellarators close to omnigeneity.
\newblock {\em Plasma Physics and Controlled Fusion}, 59(5):055014, 2017.

\bibitem{velasco2017hole}
J~L Velasco, I~Calvo, S~Satake, A~Alonso, M~Nunami, M~Yokoyama, M~Sato,
  T~Estrada, J~M Fontdecaba, M~Liniers, K~J McCarthy, F~Medina, B~Ph~Van
  Milligen, M~Ochando, F~Parra, H~Sugama, A~Zhezhera, {the LHD experimental
  team}, and {the TJ-II team}.
\newblock {Moderation of neoclassical impurity accumulation in high temperature
  plasmas of helical devices}.
\newblock {\em Nuclear Fusion}, 57(1):016016, 2017.

\bibitem{hirshman1986dkes}
S~P Hirshman, K~C Shaing, W~I van Rij, C~O Beasley, and E~C Crume.
\newblock {Plasma transport coefficients for nonsymmetric toroidal confinement
  systems}.
\newblock {\em Physics of Fluids}, 29(9):2951--2959, 1986.

\bibitem{beidler2011ICNTS}
C~D Beidler, K~Allmaier, M~Yu Isaev, S~V Kasilov, W~Kernbichler, G~O Leitold,
  H~Maa{\ss}berg, D~R Mikkelsen, S~Murakami, M~Schmidt, D~A Spong, V~Tribaldos,
  and A~Wakasa.
\newblock Benchmarking of the mono-energetic transport coefficients. results
  from the international collaboration on neoclassical transport in
  stellarators (icnts).
\newblock {\em Nuclear Fusion}, 51(7):076001, 2011.

\bibitem{maassberg1999densitycontrol}
H~Maa{\ss}berg, C~D Beidler, and E~E Simmet.
\newblock {Density control problems in large stellarators with neoclassical
  transport}.
\newblock {\em Plasma Physics and Controlled Fusion}, 41(9):1135, 1999.

\bibitem{pedrosa2015phi1}
M~A Pedrosa, J~A Alonso, J~M Garc\'ia-Rega{\~n}a, C~Hidalgo, J~L Velasco,
  I~Calvo, C~Silva, and P~Helander.
\newblock {Electrostatic potential variations along flux surfaces in
  stellarators}.
\newblock {\em Nuclear Fusion}, 55(5):052001, 2015.

\bibitem{liu2017phi1}
B~Liu, J~M Garc\'ia-Rega{\~n}a, U~Losada, J~L Velasco, J~A Alonso, H~Takahashi,
  S~Oshima, T~Wu, B~Ph~Van Milligen, C~Silva, R~Kleiber, and C~Hidalgo.
\newblock Direct experimental evidence of asymmetry in the modulation of
  potential in magnetic flux surfaces in the tj-ii stellarator.
\newblock {\em Nuclear Fusion}, submitted, 2017.

\bibitem{landreman2014sfincs}
M~Landreman, H~Smith, A~Moll\'en, and P~Helander.
\newblock {Comparison of particle trajectories and collision operators for
  collisional transport in nonaxisymmetric plasmas}.
\newblock {\em Physics of Plasmas}, 21(4):042503, 2014.

\bibitem{ida2015lhd}
K.~Ida, K.~Nagaoka, S.~Inagaki, H.~Kasahara, T.~Evans, M.~Yoshinuma, K.~Kamiya,
  S.~Ohdach, M.~Osakabe, M.~Kobayashi, S.~Sudo, K.~Itoh, T.~Akiyama, M.~Emoto,
  A.~Dinklage, X.~Du, K.~Fujii, M.~Goto, T.~Goto, M.~Hasuo, C.~Hidalgo,
  K.~Ichiguchi, A.~Ishizawa, M.~Jakubowski, G.~Kawamura, D.~Kato, S.~Morita,
  K.~Mukai, I.~Murakami, S.~Murakami, Y.~Narushima, M.~Nunami, N.~Ohno,
  N.~Pablant, S.~Sakakibara, T.~Seki, T.~Shimozuma, M.~Shoji, K.~Tanaka,
  T.~Tokuzawa, Y.~Todo, H.~Wang, M.~Yokoyama, H.~Yamada, Y.~Takeiri, T.~Mutoh,
  S.~Imagawa, T.~Mito, Y.~Nagayama, K.Y. Watanabe, N.~Ashikawa, H.~Chikaraishi,
  A.~Ejiri, M.~Furukawa, T.~Fujita, S.~Hamaguchi, H.~Igami, M.~Isobe,
  S.~Masuzaki, T.~Morisaki, G.~Motojima, K.~Nagasaki, H.~Nakano, Y.~Oya,
  C.~Suzuki, Y.~Suzuki, R.~Sakamoto, M.~Sakamoto, A.~Sanpei, H.~Takahashi,
  H.~Tsuchiya, M.~Tokitani, Y.~Ueda, Y.~Yoshimura, S.~Yamamoto, K.~Nishimura,
  H.~Sugama, T.~Yamamoto, H.~Idei, A.~Isayama, S.~Kitajima, S.~Masamune,
  K.~Shinohara, P.S. Bawankar, E.~Bernard, M.~von Berkel, H.~Funaba, X.L.
  Huang, T.~Ii, T.~Ido, K.~Ikeda, S.~Kamio, R.~Kumazawa, T.~Kobayashi, C.~Moon,
  S.~Muto, J.~Miyazawa, T.~Ming, Y.~Nakamura, S.~Nishimura, K.~Ogawa, T.~Ozaki,
  T.~Oishi, M.~Ohno, S.~Pandya, A.~Shimizu, R.~Seki, R.~Sano, K.~Saito,
  H.~Sakaue, Y.~Takemura, K.~Tsumori, N.~Tamura, H.~Tanaka, K.~Toi, B.~Wieland,
  I.~Yamada, R.~Yasuhara, H.~Zhang, O.~Kaneko, A.~Komori, and Collaborators.
\newblock Overview of transport and mhd stability study: focusing on the impact
  of magnetic field topology in the large helical device.
\newblock {\em Nuclear Fusion}, 55(10):104018, 2015.

\bibitem{sunnpedersen2016nature}
T~Sunn-Pedersen, M~Otte, S~Lazerson, P~Helander, S~Bozhenkov, C~Biedermann,
  T~Klinger, R~Wolf, H~S Bosch, and {The Wendelstein 7-X Team}.
\newblock Confirmation of the topology of the wendelstein 7-x magnetic field to
  better than 1:100,000.
\newblock {\em Nature Communications}, 7(13493), 2016.

\bibitem{klinger2017op11}
T~Klinger, A~Alonso, S~Bozhenkov, R~Burhenn, A~Dinklage, G~Fuchert, J~Geiger,
  O~Grulke, A~Langenberg, M~Hirsch, G~Kocsis, J~Knauer, A~Kr{\"a}mer-Flecken,
  H~Laqua, S~Lazerson, M~Landreman, H~Maa{\ss}berg, S~Marsen, M~Otte,
  N~Pablant, E~Pasch, K~Rahbarnia, T~Stange, T~Szepesi, H~Thomsen, P~Traverso,
  J~L Velasco, T~Wauters, G~Weir, T~Windisch, and {The Wendelstein 7-X Team}.
\newblock {Performance and properties of the first plasmas of Wendelstein 7-X}.
\newblock {\em Plasma Physics and Controlled Fusion}, 59(1):014018, 2017.

\bibitem{sanchez2015iaeap}
J~S\'anchez et~al.
\newblock {Transport, stability and plasma control studies in the TJ-II
  stellarator}.
\newblock {\em Nuclear Fusion}, 55(10):104014, 2015.

\bibitem{matsuoka2015tangential}
Seikichi Matsuoka, Shinsuke Satake, Ryutaro Kanno, and Hideo Sugama.
\newblock Effects of magnetic drift tangential to magnetic surfaces on
  neoclassical transport in non-axisymmetric plasmas.
\newblock {\em Physics of Plasmas}, 22(7):072511, 2015.

\bibitem{huang2017bootstrap}
B.~Huang, S.~Satake, R.~Kanno, H.~Sugama, and S.~Matsuoka.
\newblock Benchmark of the local drift-kinetic models for neoclassical
  transport simulation in helical plasmas.
\newblock {\em Physics of Plasmas}, 24(2):022503, 2017.

\bibitem{cary1997omni}
J~R Cary and S~G Shasharina.
\newblock {Helical Plasma Confinement Devices with Good Confinement
  Properties}.
\newblock {\em Physical Review Letters}, 78:674--677, 1997.

\bibitem{hokulsrud1987sqrtnu}
Darwin~D.?M. Ho and Russell~M. Kulsrud.
\newblock Neoclassical transport in stellarators.
\newblock {\em The Physics of Fluids}, 30(2):442--461, 1987.

\bibitem{calvo2017viena}
I~Calvo, J~L Velasco, F~I Parra, J~A Alonso, and J~M Garc{\'\i}a-Rega{\~n}a.
\newblock {Neoclassical calculation of the tangential electric field in
  stellarators close to omnigeneity (and tokamaks with broken symmetry)}.
\newblock In {\em 10th Plasma Kinetics Working Group Meeting, Wolfgang Pauli
  Institute, Vienna, 2017}, 2017.

\end{thebibliography}

\end{document}